\documentclass[preprintnumbers,twocolumn,nofootinbib,amsmath,amssymb,aps,nofootinbib]{revtex4-2} 

\usepackage{graphicx}
\usepackage{dcolumn}
\usepackage{hyperref}
\usepackage{soul}
\usepackage[T1]{fontenc} 
\usepackage{amsmath}
\usepackage{amssymb}
\usepackage{mathtools}
\usepackage{slashed}
\usepackage[section]{placeins}
\usepackage{braket}
\usepackage{upgreek}
\usepackage[bottom]{footmisc}
\usepackage[normalem]{ulem}
\usepackage{lipsum}

\newcommand{\eq}[1]{Eq.~\eqref{#1}}
\newcommand{\eqs}[2]{Eqs.~\eqref{#1} and \eqref{#2}}
\newcommand{\eqns}[3]{Eqs.~\eqref{#1}, \eqref{#2} and \eqref{#3}}

\usepackage{xcolor}

\newcommand{\e}{{\epsilon}}
\newcommand{\nn}{\nonumber} 
\newcommand{\df}{\mathrm{d}}
\def\scetg{{\rm SCET_G}}

\begin{document}

\preprint{TIFR/TH/23-7}

\title{Medium modifications to jet angularities using SCET with Glauber gluons }
\author{Ankita Budhraja}
\affiliation{Department of Theoretical Physics, 
Tata Institute of Fundamental Research, 1, Homi Bhabha Road, Colaba, 
Mumbai, 400005, India}
\email{ankita@theory.tifr.res.in}

\author{Rishi Sharma}
\affiliation{Department of Theoretical Physics, 
Tata Institute of Fundamental Research, 1, Homi Bhabha Road, Colaba, 
Mumbai, 400005, India}
\email{rishi@theory.tifr.res.in}

\author{Balbeer Singh}
\affiliation{Department of Theoretical Physics, 
Tata Institute of Fundamental Research, 1, Homi Bhabha Road, Colaba, 
Mumbai, 400005, India}
\email{balbeer@theory.tifr.res.in}

\date{\today}


\begin{abstract}
We perform a comprehensive analysis of medium modifications on ungroomed jet angularities, $\tau_a$,  within the framework of Soft-Collinear Effective Theory with Glauber gluons (SCET$_{\rm G}$). Angularities are a one-parameter family of jet substructure observables with angularity exponent $a < 2$ for infrared safety. Variation of the angularity exponent allows one to modify the relative weighting of the collinear-to-soft radiations in the jet, thereby giving access to different moments of the jet transverse momentum spectrum. In this article, we focus on $a<1$ and provide detailed results for $a=-1, 0$, and $0.5$. Within SCET$_{\rm G}$, the interactions between jet and medium constituents are mediated by off-shell Glauber gluons generated from the color sources in the medium. While medium modifications are incorporated into the jet function via the use of medium-induced splitting functions, the soft function remains unmodified for $a<1$. For all values of $a$, we find that compared to jets in vacuum, the medium-modified distributions are shifted towards smaller values of jet angularity and have a steeper fall. This redistribution of the ungroomed angularity spectrum is more apparent for a jet with a larger cone size and for higher values of $a$. We also present results for the medium sensitivity towards $p_T$ of the jet and for a jet initiated in a less central event ($10-30\%$ centrality). Finally, we provide the ratios of nucleus-nucleus and proton-proton differential angularity distributions for different angularity exponents, and for two values of the jet radius parameter.
\begin{description}
\item[Keywords]
heavy-ion Phenomenology, Jets, Glaubers, Effective Field Theory
\end{description}
\end{abstract} 

\maketitle


\section{Introduction}

Jets are complex objects that exhibit dynamics across multiple length and energy scales. Due to their sensitivity to various scales, they are useful in probing the properties of the quark-gluon plasma (QGP) created in high-energy nuclear collision experiments. Over the past decade, experiments at the Large Hadron Collider (LHC) have provided measurements of a diverse set of observables characterizing jets in heavy-ion (AA, particularly PbPb) collisions~\cite{Connors:2017ptx}.  By comparing the production cross-section in PbPb collisions, scaled by the inverse number of binary collisions, to that in proton-proton (pp) collisions, it is possible to quantify the energy loss experienced by the energetic partons as they traverse the QGP medium~\cite{ATLAS:2012tjt,ATLAS:2014ipv,ATLAS:2018gwx,ALICE:2015mjv,ALICE:2019qyj,CMS:2016uxf}.  The interactions between an energetic parton and the medium depend on the medium scales such as its temperature, size as well as other scales that appear through their interactions, such as jet quenching parameter ($\hat{q}$) or a critical angle characterizing color coherence dynamics~\cite{Baier:2000mf,Wiedemann:2009sh,Majumder:2010qh,Singh:2022dcr,Casalderrey-Solana:2012evi}. Therefore, the quenching of jets in heavy-ion collisions (HICs) is a potential probe of the emergent phenomena in many-body quantum chromodynamics (QCD) interactions in the medium. 

To fully exploit the power of jet measurements to evaluate medium properties, differential jet substructure measurements are considered useful. These measurements provide information about the distribution of particles within the jet. In recent years, there has been a growing interest in studying such observables in AA due to their potential to reveal crucial information about the microscopic properties of the QGP. For example, some of the substructure observables that have been measured so far include jet
mass~\cite{CMS:2018fof}, jet fragmentation
functions~\cite{CMS:2012nro,CMS:2014jjt}, momentum splitting
fraction~\cite{ALargeIonColliderExperiment:2021mqf}, jet shapes~\cite{CMS:2013lhm,CMS:2018jco}, and
others~\cite{ALICE:2019ykw,CMS:2020plq}.\footnote{There have been some recent studies of angularities in the medium, see Refs.~\cite{Chien:2024uax, Li:2024pfi}}
These substructure measurements provide an additional advantage in disentangling various effects -- including the initial state and cold nuclear matter (CNM) effects -- from jet-medium interactions~\cite{Chien:2015hda}.

One family of jet substructure observables of particular interest are
angularities~\cite{Berger:2003iw,Berger:2004xf}. The angularity of
a reconstructed jet in a pp (or AA) collision is defined as~\cite{Hornig:2016ahz,Larkoski:2017jix,Kang:2018qra}
\begin{equation}
\tau_a = \frac{1}{p_T}\sum_{i\in J} 
|\vec{p}_T^{\,\, i}| (\Delta {\cal R}_{iJ})^{2-a}\, ,~\label{eq:taupp} 
\end{equation} 
where $\Delta {\cal R}_{iJ} = \sqrt{(\Delta \eta_{iJ})^2+(\Delta \phi_{iJ})^2}$ and $\vec{p}_T^{\,\,i}$ denotes the transverse momentum of the particle $i$ in the jet relative to the beam axis. Here $\Delta \eta_{iJ}$ and $\Delta\phi_{iJ}$ are the rapidity and azimuthal angle separation between the particle $i$ and the jet $J$. The sum ($i\in J$) is over all particles in the reconstructed jet, and $p_T \equiv \vert \vec{p}_T\vert$
is the transverse momentum of the jet. Variation of the exponent `$a$' provides sensitivity to different sectors of the jet. For instance, when $a<0$ it probes the collinear radiation in the jet while for $a>0$ the observable also gets a significant contribution from the soft radiation. The ability to tune the angularity exponent, therefore, provides access to various moments of the jet transverse momentum ($p_T$) spectrum, which is essential for understanding the energy loss and color coherence dynamics in the medium.

The definition of angularity in $\rm pp$ collisions~\cite{Hornig:2016ahz,Larkoski:2017jix,Kang:2018qra} can be related to that in $e^{+}e^{-}$ collisions as
\begin{equation}
\tau_a^{\rm pp}=\bigg(\frac{2 E_J}{p_T} \bigg)^{2-a}\tau_a^{e^{+}e^{-}} \, ,
\label{tpp}
\end{equation}
where~\cite{Berger:2003iw,Berger:2004xf}
\begin{equation}
    \tau_a^{e^{+}e^{-}} = \frac{1}{2 E_J}\sum_{i\in J}|\vec{p}_T^{\,\, iJ}| e^{-|\eta_{iJ}|(1-a)}\, ,
\end{equation}
where $\vec{p}_T^{\,\, iJ}$ denotes the transverse momentum of the particles with respect to the jet axis and $E_J$ is the total energy of the jet. For leading-order splitting, i.e.,  $l \to q^\prime \, k$, where $l,q^\prime,k$ are momenta of corresponding particles, the expression in Eq.~\ref{tpp} can be simplified as  
\begin{align}
\tau_a^{\rm pp}
&=p_T^{2-a} \left\{x^{a-1}k_{\perp}^{2-a} +(1-x)^{a-1}q^{\prime 2-a}_{\perp} \right\}\, ,
\end{align}
where we have used $l^+=2 E_J$, with $l^+$ denoting the large light-cone momentum component of the jet parton.

The phenomenology of angularity distributions in pp collisions has been extensively studied using both Monte Carlo simulations~\cite{Caletti:2021oor,Reichelt:2021svh} and effective field theories (EFTs) such as Soft Collinear Effective Theories (SCET)~\cite{Kang:2018qra,Kang:2018vgn}, and results have been compared with the measured distributions~\cite{ALICE:2021njq, Kang:2018qra, Reichelt:2021svh}. 

The main goal of this paper is to qualitatively understand medium modifications of the jet angularity distributions for a range of angularity exponents, and to quantify the effect of medium-induced energy loss within the leading opacity expansion of SCET-based approach, by considering single and multi-gluon emissions. We also incorporate initial state CNM effects.  
There could be other competing effects affecting the distribution of angularities in the medium, such as color (de)coherence. These effects cannot be captured at the leading medium-induced emissions and their systematic inclusion requires higher-order loop calculations (next-to-next-leading order) as pointed out in Ref.~\cite{Mehtar-Tani:2024smp}, which so far have not yet been computed. While we consider a simple Bjorken expansion of the medium, a more realistic hydrodynamic evolution of the medium may also play a role. Despite these simplifications, our calculation includes the key dynamics of medium-induced Bremsstrahlung of energetic partons. 

In the literature, numerous approaches~\cite{Arnold:2002zm,Caucal:2018dla,Gyulassy:2000er,Guo:2000nz,Wiedemann:2000za,Barata:2021wuf} have been developed to model the propagation of energetic partons within the QGP in a qualitative manner. In this article, we will follow the method developed in Refs.~\cite{Ovanesyan:2011xy,Ovanesyan:2011kn} to achieve a systematic resummation of the relevant logarithms that arise in the fixed-order perturbative distributions. This approach builds upon SCET and expands it to incorporate interactions between collinear partons in a jet and the thermal constituents of the medium, mediated by Glauber gluons induced by background color sources in the medium.
In the absence of the medium, SCET can be used to write down the differential cross-section for angularities in a factorized form as a
convolution of the parton distribution functions with the hard function describing initial state production via hard scatterings, a jet function describing the evolution of the collinear modes of the jet, and a soft function that incorporates dynamics of soft emissions. With this factorization, one can arrive at an operator definition of the EFT modes, which can be used to systematically compute fixed-order jet and soft functions at any perturbative order. Additionally, it was shown in Ref.~\cite{Ritzmann:2014mka} that the same fixed-order jet function can be obtained using the QCD splitting functions by integrating them over appropriate collinear phase-space factors. For the case of generalized angularities, we show this explicitly in Appendix~\ref{app:Jet}. 

Assuming a similar factorization holds in the presence of the medium as well, the identification of the jet function with the splitting functions becomes important, as a possible medium effect can be incorporated through medium-induced splittings. 
A similar approach for other observables has been used in Refs.~\cite{Chien:2015hda,Chien:2016led,Li:2019dre}. 
Additionally, due to the interactions between jet partons and medium constituents, an amount of energy is lost by the jet to the medium. 
For reconstructed jets with radius $R$, this
energy loss can be related to the out-of-cone jet radiation.  
We model this effect by a fractional shift in $p_T$ of the jet- producing parton, by an amount $\varepsilon$, computed for a single gluon emitted out-of-the-jet-cone. We also consider the effect of fluctuations in the emitted number of gluons~\cite{Gyulassy:2001nm}. This energy loss, in addition to the medium-induced splittings, leads to a shift of the distribution to smaller angularity values. We find that the shift is more apparent for small $p_T$ jet distributions. 
 
In this paper, we present new analytic results on angularity distributions for fully reconstructed jets in pp and PbPb collisions at $\sqrt{s} = 5.02\, \rm{TeV}$. Our results are presented at next-to-leading-logarithmic (NLL$^{\prime}$) accuracy, where the primed counting implies that the fixed-order coefficients at ${\cal{O}}(\alpha_s)$ are also included. Specifically, we present results for jets reconstructed with $R = 0.2$ and $R = 0.4$ in two ranges of jet transverse momenta, $40<p_T<60\, \rm{GeV}$ and $80<p_T<100\, \rm{GeV}$. PbPb results are presented for two different centrality bins, i.e., 0-10\% and 10-30\%. The ratios of angularity distributions between PbPb and pp are presented for three values of the exponent $a= -1, 0, 0.5$.

The rest of the paper is organized as follows. In Sec.~\ref{sec:SCETG} we briefly review the SCET$_{\rm G}$ formalism.  In Sec.~\ref{sec:factorization} we discuss the factorization theorem and introduce the relevant definitions. In Sec.~\ref{sec:angularitiespp} we present a brief summary of the resummed pp angularity results. In Sec.~\ref{sec:angularitiesAA} we compute the medium-modified jet function and provide the relevant medium parameters. Further, we provide the results for angularity distributions for pp and AA collisions in Sec.~\ref{sec:results}. Readers interested in the final results can go directly to this section and look at Figure~\ref{fig:compare_R}, Figure~\ref{fig:compare_pT}, and Figure~\ref{fig:compare_a}. In Sec.~\ref{sec:discussion}, we summarize the results. We have moved some details to the appendices. In Appendix~\ref{app:PlusDistributions}, we provide some details about the plus distributions. In Appendix~\ref{app:pert}, we provide explicit details of the perturbative calculations.  


\section{Soft-Collinear Effective Theory with Glauber gluons}
\label{sec:SCETG}

SCET~\cite{Bauer:2000ew,Bauer:2002uv,Bauer:2000yr,Bauer:2001ct,Bauer:2001yt,Bauer:2002nz} is an effective field theory of QCD designed to separate scales in high-energy scattering processes to describe the dynamics of energetic quarks and gluons. It has been successfully applied to study jet production, its subsequent evolution, and to compute a large class of jet substructure observables up to the state-of-the-art next-to-next-to-leading-logarithmic (NNLL) accuracy (see~\cite{Larkoski:2017jix,Kogler:2018hem} and references therein). Before we discuss the EFT describing jet-medium interactions, we outline the necessary ingredients of SCET.

In general, the relevant degrees of freedom described by SCET are the collinear and soft modes. Partons that are highly boosted along a specific direction constitute the collinear modes. On the other hand, the momentum components of soft modes have a homogeneous scaling. The large momentum component along the collinear direction makes the use of light-cone coordinates more convenient to describe the EFT modes. In these coordinates, any four-momentum can be expressed as  $p^{\mu}\equiv(p^+,p^-,p_{\perp})$, with $p^+=n\cdot p$ and $p^- = \bar{n}\cdot p$, where
$n^{\mu},\bar{n}^{\mu}$ are light-like vectors such that $n^2=\bar{n}^2=0$ and $n\cdot\bar{n}=2$. For a collinear parton along $\Bar{n}$ direction, we have $p^+\gg p_{\perp}\gg p^-$ hierarchy while, for a soft parton, there is no hierarchy in momentum components. 
More explicitly, for {\it angularities}, in terms of the power counting parameter $\lambda\ll 1$ (corresponding to $\lambda \sim \tau_a^{\frac{1}{2-a}}$), the momentum scaling of the SCET modes is as follows~\cite{Budhraja:2019mcz}
\begin{align}
 p_n\sim Q(\lambda^2,1,\lambda)\, ,  \,\,
 p_s \sim Q(\lambda^{2-a},\lambda^{2-a},\lambda^{2-a})   \, ,
\end{align}
where $Q$ is the scale of the hard interaction, typically of the order of $p_T$ for pp (and AA) collisions. The $\bar{n}$ collinear mode has a scaling analogous to $n$ collinear mode, with the $p^+$ and $p^-$  components interchanged.

For the case of $a<1$ angularities, the soft mode always scales smaller than ${\cal O}(\lambda)$, and hence $p_s^2 \ll p_n^2$. In the literature, such soft modes are usually known as {\it ultrasoft} modes and the corresponding EFT with these degrees of freedom as SCET$_{\rm I}$. However, for convenience, we will refer to them  as {\it soft} modes. The gauge-invariant collinear quark and gluon fields are built out of the SCET collinear fields ($\xi_n$,$A_n$) and the collinear Wilson line ($U_n$). At leading order in $\lambda$, the interaction term between soft gluons and the collinear fields in the SCET Lagrangian is decoupled through a BPS field redefinition~\cite{Bauer:2001yt}. 

To describe jet dynamics in a nuclear medium, additional degrees of freedom are required that account for interactions between the jet and the medium constituents. In the literature, these interactions 
have been modeled by incorporating additional modes that can be thought to originate from the static background color gluon fields of the medium and are known as the `Glauber modes'.  
The corresponding EFT with such extra degrees of freedom is known as SCET$_\text{G}$~\cite{Idilbi:2008vm,DEramo:2010wup,Ovanesyan:2011xy}. Glaubers are non-propagating off-shell degrees of freedom, with transverse momentum much larger than their other light-cone momentum components, $k_{\perp}^2 \gg k^{+}k^{-}$. 
The relevant characteristic scale for the Glauber modes can be of the order of the medium temperature $T$ or the Debye screening mass $m_D$. Assuming that both the jet and the medium have the same virtuality, we then require
\begin{equation}
    \tilde{\lambda} \sim \frac{T}{Q} \lesssim \lambda \sim \tau_a^{\frac{1}{2-a}}  \ll 1\, ,
\end{equation}
to allow for the interactions between jet partons and the medium constituents. From here onward, we will not distinguish between the scales $\lambda$ and $\tilde{\lambda}$ and refer to both simply as $\lambda$.  In an alternative SCET-based approach~\cite{Vaidya:2020cyi},  factorization has been derived for the case where the virtuality of the collinear mode is higher than the medium modes ($\tilde\lambda \ll \lambda$), for jet observables such as the energy-energy correlator and inclusive jet production in Refs.~\cite{Mehtar-Tani:2024smp,Singh:2024vwb,Singh:2024vwb}. In this approach, the production of high virtuality modes is described by a matching function. However, in this framework, the factorization has not yet been derived for observables such as {\it jet angularities} that necessarily require soft dynamics. Moreover, the computation is so far only limited to quark jets.

 In the present study, we follow Ref.~\cite{Ovanesyan:2011xy}, which has been used to obtain the relevant medium quantities such as the medium-induced splitting functions which we use in our computation. 


\section{Factorization Theorem for Angularities}
\label{sec:factorization}

The factorization theorem for angularities with $a<1$ was first proved for $e^+\,e^-$ collisions in Ref.~\cite{Hornig:2009vb} and subsequently studied in pp collisions in Ref.~\cite{Kang:2018qra}.\footnote{When $a$ is close to 1, the soft and collinear modes have similar offshellness, i.e. $p_s^2 \sim p_n^2$. Such modes are described by SCET$_{\rm II}$ framework and are beyond the scope of this work.} For AA collisions, the typical scale of the jet-producing parton is usually much larger than the medium scales; therefore, it is natural to assume that the jet first undergoes a vacuum-like emission, and as the virtuality goes down, interactions with the medium become relevant. Therefore, at leading power, we write the following form of the factorization theorem for medium-modified angularity distributions
\begin{align}
&\frac{d\sigma^{{\rm AA}\to ({\rm jet} \, \tau_a) {\rm X}}}{d\tau_a\, dp_T\, d\eta} \!=\! \frac{2p_T}{s} \!\sum_{abci}  \int \!d\varepsilon\, P_c(\varepsilon) \!\! \int_{x_a^{\rm min}}^1 \!\! \frac{d x_a}{x_a} \tilde{f}_a(x_a,\mu) \!  \nonumber \\
& \int_{x_b^{\rm min}}^1 \! \frac{d x_b}{x_b} 
\tilde{f}_b(x_b,\mu) 
\int_{z^{\rm min}}^1 \frac{d z}{z^2} 
H_{ab}^{c}(x_a,x_b,\hat{s},\hat{\eta},\tilde{p}_{T}/z,\mu)  \nonumber \\
& {\cal H}_{c\to i}(z,\tilde{p}_{T}\, R,\mu) {\cal J}(\tau_a^n,\tilde{p}_{T},\mu) \otimes {\cal S}(\tau_a^s,\tilde{p}_{T},R,\mu) \, ,
\label{eq:AAfact}
\end{align} 
which is valid up to $\mathcal{O}(R^2, \tau_a^{\frac{1}{2-a}}/R)$. Here, $\tilde{f}$ denotes the nuclear parton distribution functions (nPDFs)~\cite{PhysRevD.80.094004,Kovarik:2015cma,Vitev:2006bi,Vitev:2007ve} and $\sqrt{s}$ is the hadronic center of mass energy. $P_c(\varepsilon)$ is the distribution of the fractional energy loss experienced by a jet initiated by the parton $c$ due to medium-induced out-of-cone radiation, and $\tilde{p}_{T} = p_T(1+\varepsilon)$. Details of $P_c$ and $\tilde{f}$ are given in Section~\ref{sec:elossdetails}.  Moreover, $H_{ab}^{c}$ denotes the hard scattering of partons $a$ and $b$ to produce the final state parton $c$, which is observable-independent and is the same as in vacuum. To next-to-leading order (NLO), this has been worked out explicitly in Refs.~\cite{Aversa:1988vb,Jager:2002xm}. Note that the hard function is expressed in terms of the partonic variables, i.e. the partonic center of mass energy $\hat{s}=x_a x_b s$, transverse momentum $\hat{p}_T= \tilde{p}_T/z$, and 
rapidity $\hat{\eta} = \eta - \frac{1}{2} \log(x_a/x_b)$. We further define new variables $V$ and $W$ in terms of the hadronic variables~\cite{Aversa:1988vb,Kaufmann:2015hma}
\begin{equation}
    V = 1-\frac{\tilde{p}_T}{\sqrt{s}}e^{-\eta} \, , \hspace{0.5cm} W = \frac{\tilde{p}_T^2}{s V (1-V)} \, .
\end{equation}
In terms of these variables, the lower limits of the integrals in $x_{a,b}$ and $z$ can be expressed as
\begin{equation}
    x_a^{\rm min} = W\, , \hspace{0.1cm} x_b^{\rm min} = \frac{1-V}{1-V W/x_a} \, , \hspace{0.1cm} z^{\rm min} = \frac{1-V}{x_b} + \frac{V W}{x_a}\, .
\end{equation}
The other ingredients appearing in the factorization theorem of \eq{eq:AAfact} are the hard matching function ${\cal H}_{c\to i}$, and the observable-dependent jet and the soft functions, ${\cal J}$ and $\cal S$. The hard matching function, ${\cal H}_{c\to i}$, gives the matching coefficient for the transition from the parton generated at hard scattering, $c$, to the jet initiating parton $i$. This includes the effect of parton $c$ radiating any number of partons that are not contained inside the jet of radius $R$ and resums the logarithms of jet radius R through a DGLAP evolution equation, as discussed in Appendix~\ref{sec:matching} while the logarithms of the measurement functions are resummed through the anomalous dimensions corresponding to the jet and the soft function.  The matching function has been computed to the desired NLO accuracy in Ref.~\cite{Kang:2016ehg}.

The jet function ${\cal J}$ appearing in \eq{eq:AAfact} is modified  by the jet-medium interactions and can be written as ${\cal J} = {\cal J}^{\rm vac}+{\cal J}^{\rm med}$, where ${\cal J}^{\rm med}$ describes medium-induced collinear emissions. We compute this medium-modified contribution through the use of medium-induced splittings computed in Ref.~\cite{Ovanesyan:2011xy} and show that at the order we are working at, the medium modifications enter through the modified initial conditions for the renormalization group (RG) equations.

It is well known that the interaction between the energetic parton and the medium is dominated by forward scatterings that scale as $1/\theta^4$, where $\theta$ is angular deflection of energetic parton from its original direction after interacting with the medium~\cite{Ovanesyan:2011xy,Vaidya:2020cyi}. In full theory, the dominant contributions are calculated in Refs.~\cite{Gyulassy:1999zd,Arnold:2002zm}. In the EFT picture, these dominant interactions are precisely described by the jet function~\cite{Ovanesyan:2011xy}. 
Further, for $a<1$  the soft mode of the jet scales as $Q(\lambda^{2-a},\lambda^{2-a},\lambda^{2-a})$ and has a virtuality
$\sim Q^2\lambda^{4-2a}$ which is smaller than the virtuality of the Glaubers which is $\sim Q^2\lambda^2$. Coupling of the soft mode with the Glauber gluons will increase its virtuality, driving it offshell. Although these interactions can in principle be factored out in terms of the Wilson line, but it requires further investigation and a systematic incorporation of soft-Glauber interactions in the SCET$_{\rm G}$ framework is an open problem. For our case, we do not consider such couplings and keep only vacuum contributions in the soft function as explained above. 
 
Finally, $\tau_a^{n,s}$ is the angularity in the collinear and soft sectors, and the total jet angularity $\tau_a$ is given by the sum of the two, $\tau_a = \tau_a^n + \tau_a^s$. The symbol $\otimes$ in the final line of \eq{eq:AAfact} represents convolutions over the measurement observable $\tau_a$. Note that, in principle, the jet function will also be a function of the jet radius $R$.  However, it was shown in Ref.~\cite{Ellis:2010rwa} that the leading power divergent contribution to the jet function does not depend on the jet boundary $R$ to NLO accuracy. Therefore, we suppress this dependence in the jet function.


\section{Jet Angularity Distributions in vacuum}
\label{sec:angularitiespp}

In this section, we briefly summarize the resummed angularity distributions for the vacuum, incorporate the leading non-perturbative effects, and compare to experimental data~\cite{ALICE:2021njq}.

\subsection{Jet function}

It has been shown in Ref.~\cite{Ritzmann:2014mka} that the jet functions in SCET can be directly computed from the spin-averaged QCD splitting functions with appropriate phase space factors. This averts the use of the more complicated collinear Feynman rules of SCET. 
At the lowest order (i.e., $l \to k\, q^\prime$), the jet function for generalized angularities 
can be expressed as   
\begin{align}
&\mathcal{J}^{\rm vac}_{i} \equiv \mathcal{J}_{i}(\tau_a,p_T,R,\mu)=\frac{\alpha_s(\mu)}{\pi}\frac{e^{\epsilon \gamma_E}\mu^{2\epsilon}}{\Gamma(1-\epsilon)} \sum_{j, k}\int dx \frac{dk_{\perp}}{k_{\perp}^{2\epsilon-1}} \nonumber \\
&\qquad \qquad \times {\cal{P}}_{i \to jk}(x,k_{\perp})\delta(\tau_a-\hat{\tau}_a)\, ,
\label{jfun}
\end{align}
where $x$ is the momentum fraction carried by the final state parton and $k_\perp$ is its transverse momentum. Here $i, j, k$ are the indices for the initial and final state partons, and $l, k, q^\prime$ are their respective momenta. In the above equation, $\epsilon$ is the usual dimensional regulator and $\mu$ is the associated scale for factorizing hard, collinear, and soft modes of the theory. Note that for the vacuum jet function, $\tilde{p}_T = p_T$, hence we use this notation for this part of the discussion. 

For the vacuum case, the transverse momentum dependence trivially factors out from the splitting functions as
$\mathcal{P}_{i \to jk}(x,k_{\perp})=\mathcal{P}_{i \to jk}(x)/k_{\perp}^2$,
with $\mathcal{P}_{i \to jk}(x)$ being the usual Altarelli-Parisi QCD splitting functions.   
The fixed order jet functions can then be obtained from the knowledge of the splitting functions and we have verified that they agree exactly with the ones computed from the standard operator definition, see Appendix~\ref{app:Jet}~\cite{Kang:2018qra}. 
At NLO, the jet anomalous dimension is given as
\begin{align}
   \gamma_{{\cal J}_i}(\tau_a,p_T,R,\mu)&= \bar{\alpha}_s\bigg\{\delta(\tau_a) \Big(2 b_i +\frac{(2-a)}{(1-a)}\, L_n\, C_i\Big) \nonumber \\
   &\qquad - \frac{2}{1-a} C_i \bigg[\frac{1}{\tau_a}\bigg]_{+} \bigg\} \, ,
   \label{eq:janomal}
\end{align}
where
$$ 
C_i \, , \, b_i = 
\begin{cases}
    C_F \, , \, \frac{3}{4}\, C_F \hspace{0.7cm}: &\quad {\text{for quark jet}}\, , \\
    C_A \, , \, \frac{1}{4}\,\beta_0 \hspace{0.85cm}: &\quad {\text{for gluon jet}}\, .
\end{cases} 
$$
Here $\bar{\alpha}_s \equiv \alpha_s(\mu)/\pi$, $L_n \equiv \ln(\mu^2/p_T^2)$ and $\beta_0 = (\frac{11}{3}C_A-\frac{4}{3}T_R\, N_f)$.
 Note that the jet anomalous dimension contains $a$-dependent pieces which appear to diverge as $a$ approaches 1 since the recoil of the collinear parton becomes an ${\cal{O}}(1)$ effect.

\subsection{Soft Function}
Unlike the jet function, the soft function gets a significant leading contribution from the jet radius parameter, $R$. 
For the k$_{\rm t}$-type jet algorithms~\cite{Catani:1993hr,Cacciari:2008gp,CMS-PAS-JME-09-001}, this amounts to the restriction that the relative angle between any two pair of partons should be less than $R$~\cite{Ellis:2010rwa}, i.e., $ \Theta_{\rm alg} \equiv \Theta\!\left(\omega x\, (1-x) \tan\frac{R}{2}-k_{\perp}\right)\, $.
For the case of a central emission ($\eta = 0$), $\omega = 2\, p_T$. 
The soft function 
has been computed previously in Ref.~\cite{Ellis:2010rwa} 
and its anomalous dimensions are given as, 
\begin{equation}
\gamma_{{\cal S}_i}(\tau_a,p_T,R,\mu) = \frac{2\bar{\alpha}_s\, C_i}{(1-a)} \bigg\{\left[\frac{1}{\tau_a}\right]_{+}\!\!\!-L_s\, \delta(\tau_a)\bigg\} \, ,
\label{eq:sanomal}
\end{equation}
where $L_s \equiv \ln(\mu\, R^{1-a}/p_T)$ is the large logarithm associated with the soft scale. The final ingredient of our factorization theorem is the hard matching function ${\cal H}_{c\to i}$ which is observable-independent and has been computed in Ref.~\cite{Kang:2016ehg}. 

\vspace{5pt}

\subsection{Differential angularity spectrum}

\begin{figure*}[t]
\centering
\hspace{-0.35cm}\includegraphics[scale=0.6]{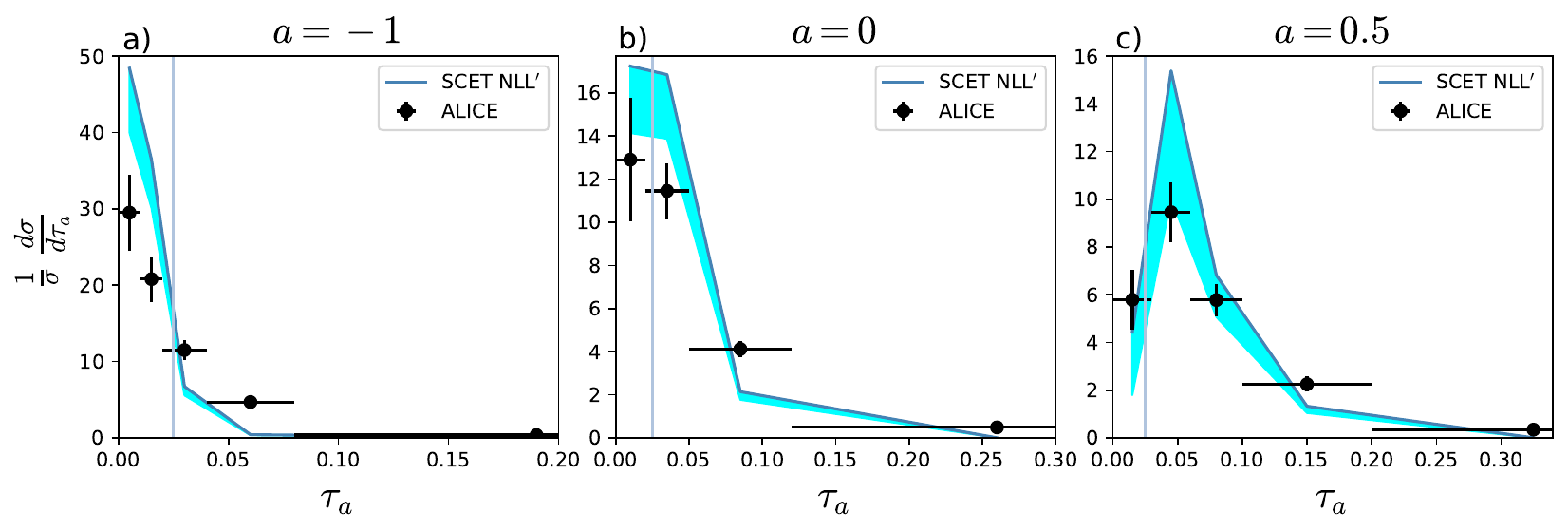}
\caption{Comparison of ungroomed jet angularity distributions $\tau_a$ in pp collisions, as reported by ALICE~\cite{ALICE:2021njq} for $R=0.4$ in the range $80<p_T<100\, \rm{GeV}$ against our analytical NLL$^{\prime}$ results convoluted with shape function, for various $a$ values. The solid black points represent the ALICE data and error bars on these include the statistical and systematic uncertainties, which we have added in quadrature. The solid blue curve corresponds to our theoretical calculation obtained by taking $\mu$ to be the canonical value, $p_T$. The shaded cyan region shows the theory error band whose lower boundary is obtained by setting $\mu=2\,p_T$. The theoretical band corresponding to using $\mu = p_T/2$ is roughly symmetric about the blue line and is not shown here for clarity. The exponent $\alpha$ used in the ALICE study is related to the standard angularity exponent that we have used by $\alpha=2-a$. {$\sigma$ in the figure is the integrated cross-section which can be obtained from Eq.~\ref{eq:convolutionNP} as discussed above. The theoretical distributions are normalized so that the area under the central value of $\mu$ distribution is unity. This is consistent with the normalization used in the ALICE data
~\cite{ALICE:2021njq}}.} 
\label{fig:pp_compare}
\end{figure*}

In the physical space of the observable, the RGEs are written as convoluted integro-differential equations. In order to solve these integro-differential evolution equations, we transform to the Fourier space so that the convolutions are turned into multiplicative RGEs (see Appendix~\ref{app:fourier} for details). 
Transforming back to the physical space, the resummed angularity jet and soft functions read as
\begin{align}
\hspace{-0.3em}{\cal J}_i^{{\rm NLL^{'}}}\!\!\!(\tau_a; \mu) &\!=\! \frac{e^{K_{\gamma_J}^i \!\!+2 K_{\Gamma_{\!J}}^i \!\!+\frac{2\eta_J^i}{j_J}}}{\Gamma(\frac{-2\eta_J^i}{j_J}) \, \tau_a^{1+\frac{2\eta_J^i}{j_J}}} \frac{\mu_J^{2\eta_J^i}}{p_T^{2\eta_J^i}} \!\!\left(\!1\!+\!\frac{\bar{\alpha}_s\, f_i^J(a)}{2-a}\!\right) \label{eq:Jsol} \!, \\
\hspace{-0.35em}{\cal S}_i^{\rm NLL^{'}}\!\!\!(\tau_a;\mu) &\!=\!\! \frac{e^{-2 K_{\Gamma_{\!S}}^i\!\! -\frac{2\eta_S^i}{j_S}}}{\Gamma(\frac{2\eta_S^i}{j_S})\, \tau_a^{1-\frac{2\eta_S^i}{j_S}}} \frac{\tilde{\mu}_S^{-2\eta_S^i}}{p_T^{-2\eta_S^i}} \!\! \left(\!1\!+\!\frac{\bar{\alpha}_s\, f_i^S(a)}{2-a}\!\right)\! ,
 \label{eq:Ssol}
\end{align}
where $\tilde{\mu}_S \equiv \mu_S\, R^{1-a}$. Here $K_{\Gamma}$, $K_\gamma$ and $\eta_{\Gamma}$  are the evolution kernels that encode the running of the anomalous dimensions between various scales and for brevity, we have suppressed their scale dependence. Their explicit expressions and expansion up to NLL order can be found in Appendix~\ref{app:pert}. Finally, the constants $f_i^{J,S}$ are the fixed order one-loop contributions of the jet and soft functions, respectively. Their explicit form can be obtained from the one-loop results by setting all logarithms to their natural scale. 
The resummed jet and soft functions are evolved from their natural scales $\mu_J \sim p_T \tau_a^{1/(2-a)}$, $\mu_S \sim p_T\, \tau_a\, R^{a-1}$ to the scale of the hard matching function $\mu_{\cal H} \sim p_T R$ which is then ultimately run up to the hard scale $\mu_H \sim p_T$. 

With all the necessary pieces from \eqns{eq:Jsol}{eq:Ssol}{eq:match}, we obtain the perturbative cross-section for angularities in the vacuum. The perturbative angularity spectrum in the small-$\tau_a$ is also sensitive to hadronization effects. We incorporate this by implementing a shape function.

\subsection{Non-perturbative Effects to Soft Function}
The relevant region of the observable where non-perturbative effects become important can be identified by demanding $\mu_S \simeq \Lambda_{\rm QCD}$, i.e.  
$\tau_a\approx \Lambda_{\rm QCD}/(p_T\, R^{a-1})$. 
For consistency with the recently published ALICE analysis~\cite{ALICE:2021njq}, we define the region of non-perturbative dynamics for ungroomed angularities to be $\tau_a \approx \Lambda_{\rm QCD}/(p_T\, R)$, independent of the exponent `$a$'. 
It has been shown for SCET$_{\rm I}$-type observables that the total cross-section corrected for non-perturbative effects can be obtained by convolving the resummed perturbative distribution with an appropriately chosen shape function, ${\cal S}_{\rm np}$ as ~\cite{Stewart:2014nna}
\begin{align}\label{eq:convolutionNP}
\hspace*{-0.5cm}\frac{d\sigma}{d\eta dp_T d\tau_a}\, =\int dk\, \frac{d\sigma^{\rm pert}}{d\eta dp_Td\tau_a}\left(\tau_a - \frac{k}{p_T R} \right)\, {\cal S}_{\rm np}(k) \,.
\end{align}

From here, we can obtain $ {d\sigma}/{d\tau_a}$ by integrating over the specified $ p_T$ range and consider central events with rapidity $ \eta = 0 $. The corresponding integrated cross-section is then given by
$
\sigma = \int d\tau_a \, \frac{d\sigma}{d\tau_a}.
$
For the non-perturbative shape function, we use the following form~\cite{Aschenauer:2019uex}
\begin{equation}\label{eq:Fk}
{\cal S}_{\rm np}(k)=\frac{4k}{\Omega_a^2}\exp\Big(\!-\frac{2k}{\Omega_a}\Big)\,,
\end{equation}
which is normalized to unity and has the first moment as $\Omega_a$. Note that the shape function only relies on a single parameter, $\Omega_a$. 
Furthermore, for angularities, it was argued that the `$a$' dependence factors out as
$\Omega_a=\Omega_{a=0}/(1-a)$
~\cite{PhysRevD.75.014022}. The factor $\Omega_{0} \equiv \Omega_{a=0}$ can be universally estimated from a global fit to the jet angularity data for different choices of `$a$' with the demand to provide an appropriate description in the perturbative region of the distribution.

In Figure~\ref{fig:pp_compare}, we qualitatively compare our theoretical results with ALICE results~\cite{ALICE:2021njq} for a jet of radius $R=0.4$ and $80<p_T<100$\,\rm{GeV} in pp collisions. The distributions are normalized to unity. The theoretical error band is obtained by varying the scale $\mu$ from $p_T/2$ to $2\,p_T$ and for clarity we have shown only the lower error band in this figure. We note that we capture the trend seen in the experimental data and we have checked the same for other jet parameters as well, i.e., $R=0.2$ and $40<p_T<60$\,\rm{GeV}. However, we emphasize the fact that a quantitative comparison will require a careful analysis and inclusion of other factors such as the profile scales, non-global logarithms (NGLs), as well the effect of charged particle reconstruction. In this study, we do not consider these effects as our goal is to consistently (within the EFT) understand medium modifications on jet angularity, which are presented in Sec.~\ref{sec:MediumJetFunc}.

The vertical solid line, in Figure~\ref{fig:pp_compare}, represents the boundary below which the shape and magnitude of the distribution is completely determined by the non-perturbative effects. Fixing $\Omega_0$ such that we obtain a qualitative description of the experimental data for all the $a$ values in the perturbative regime, gives the value $\Omega_0 = 0.35$ \rm{GeV}  for a jet with $80<p_T<100\, \rm{GeV}$ and $R=0.4$.

Having established the consistency of our theoretical calculations, we now discuss the jet-medium interactions within the SCET$_{\rm G}$ framework that forms the main objective of our study in the next section.


\section{Jet angularity Distributions in the medium}
\label{sec:angularitiesAA}
In this section, we first discuss the medium-modified angularity jet function obtained from the medium-induced splitting functions. Next, we discuss the assumptions about the QGP medium that we use for the evaluation of medium parameters, such as temperature ($T$), system size ($L$), and mean free path ($\lambda_g$). These quantities are relevant for medium-induced splittings. We show these parameters for two centrality bins, i.e., $0-10$\% and $10-30$\%, obtained using the Glauber model~\cite{Miller:2007ri}. Finally, we present our results for the medium-modified angularity distributions and their ratios with the pp baseline (PbPb/pp) for various jet parameters.

\subsection{Medium-modified Jet Function}
\label{sec:MediumJetFunc}
For a given initial energetic parton ($q$ or $g$) that initiates the jet, we evaluate the medium-modified jet function which can be obtained by formally replacing the vacuum splitting function ${\cal{P}}(x,k_\perp)$ in Eq.~\ref{jfun} by the appropriate splitting functions in the medium (see details below Eq.~\ref{eq:jmedphase}). The medium-induced splitting functions have been calculated in various frameworks~\cite{Armesto:2009ab,Ovanesyan:2011kn,Ovanesyan:2011xy,Mehtar-Tani:2019ygg,Barata:2021wuf,Attems:2022ubu}. We will follow the basic picture of Gyulassy, Levai and Vitev (GLV)~\cite{Gyulassy:1999zd,Gyulassy:2000er,PhysRevLett.85.5535} approach in which the medium is modeled by a collection of static scattering centers. As the parton traverses the medium, it exchanges momentum with these centers, which triggers medium-induced radiations. The splitting probability can be written as a series in the number of medium scatterings -- the opacity expansion~\cite{Gyulassy:2000er,PhysRevLett.85.5535}. The mean number of scatterings in a uniform medium is $n=L/\lambda_{g/q}$, where $L$ is the length of the medium traversed by the jet, and $\lambda_{g/q}$ is its mean free path. If $n\sim 1$, i.e., the medium is dilute, it is sufficient to keep the first few terms in the opacity expansion.

In particular, the medium-induced splitting functions were calculated at leading order in the opacity expansion (i.e., single scattering with medium) in Refs.~\cite{Ovanesyan:2011kn,Ovanesyan:2011xy}, in the $\scetg$ framework. We will use these to calculate the medium-modified angularity jet function. For recent work on higher-order opacity contributions, see Ref.~\cite{Sievert:2019cwq}. At NLO, the medium-modified jet function is given by
\begin{equation}
\begin{split}
&{\cal J}_i^{\rm med}(\tau_a,\varepsilon)=\int d\Phi_2^c \; \sigma_2^c \,\, \Theta_{\rm alg}
\\ 
&\qquad \delta\Bigl(\tau_a-\tilde{p}_T^{2-a} \bigl[x^{a-1}k_{\perp}^{2-a} 
+(1-x)^{a-1}q_{\perp}^{\prime 2-a} \bigr]\Bigr)\,,
\label{eq:jmedphase}
\end{split}
\end{equation}
where $\Theta_{\rm alg}$ sets jet algorithm constraint. Here, the $\delta$ function imposes the final-state value of the measured $\tau_a$, and $\tilde{p}_T = p_T(1+\varepsilon)$ (see Eq.~\ref{eq:AAfact}) accounts for the energy loss. For compactness, we have suppressed the dependence of the jet function on the scales $p_T$ and $\mu$.  In Eq.~\ref{eq:jmedphase}, the two-particle phase space $d$-dimensions, $d\Phi_2^c$ is given by
\begin{align}
d\Phi_2^c &= 2(2\pi)^{3-2\epsilon} (2E_J) \! \int \! \frac{d^d k}{(2\pi)^{d-1}} \delta(k^2) \! \int \! \frac{d^d {q^\prime}}{(2\pi)^{d-1}}\nonumber\\
&\delta({q^\prime}^2) \delta(2E_J-k^+-{q^\prime}^+) \delta^{d-2}(k_{\perp}+{q^\prime}_{\perp}) \;,
\label{eq:phase}
\end{align}
The last two $\delta$ functions in \eq{eq:phase} represent energy and transverse momentum conservation and the other two $\delta$ functions represent the on-shell conditions for the two final state partons. 

Finally, the two-particle cross-section, $\sigma_2^c$, is given in terms of the splitting function in the medium by the relation,
\begin{equation}
\sigma_2^c = \sum_{j,k}\left(\frac{\mu^2\, e^{\gamma_E}}{4\pi}\right)^{\!\!\epsilon} 2 g_s^2 \,{\cal P}^{\rm med}_{i \to jk}(x, k_{\perp})\, ,
\label{eq:sig}
\end{equation}
where $g_s$ is the coupling between Glauber gluon and jet parton and ${\cal P}^{\rm med}_{i \to jk}(x, k_{\perp})$ are the medium-induced splittings with $x$ being energy fraction carried by the radiated gluon. The summation over the indices $j, k$ indicates that one needs to sum over all possible splittings for a given jet-initiating parton $i$.

Note that in \eq{eq:jmedphase}, both the splitting functions in $\sigma_2^c$ and the transverse momentum of the partons $k_{\perp}, {q^\prime}_{\perp}$ are modified due to Glauber insertions from the medium.

At leading power, the jet radius contribution is power suppressed in the relevant kinematical regime of interest, i.e. $\tau^{1/2-a} \ll R$. Therefore, we set $\Theta_{\rm alg}\to 1$.  Simplifying the phase-space integrals by using the $\delta$ functions, we obtain a simpler form for the angularity jet function,
\begin{align}
&{\cal J}_i^{\rm med}(\tau_a,\varepsilon) = \frac{\alpha_s\mu^{2\epsilon} e^{\epsilon \gamma_E}}{\Gamma(1-\epsilon)} \sum_{j,k} \int dx\, \frac{dk_{\perp}}{k_{\perp}^{2\epsilon-1}} {\cal P}_{i \to jk}^{\rm med}(x,k_{\perp})  \nonumber \\
&\quad \times\delta\left(\tau_a-\tilde{p}_T^{a-2} k_{\perp}^{2-a}(x^{a-1}+(1-x)^{a-1})\right)\;.
\label{eq:Jmedwithdelta}
\end{align}
Here $\alpha_s=g_s^2/4\pi$ is the coupling strength with the medium, which we take as constant over the temperature range of the evolving medium. Finally, carrying out the remaining $k_{\perp}$ integration using the measurement $\delta$ function, the medium-modified angularity jet function acquires the form
\begin{widetext}
\begin{align}
 {\cal J}_i^{\rm med}(\tau_a,\varepsilon) &= \frac{\alpha_s}{\vert 2-a\vert} \frac{\mu^{2\epsilon} e^{\epsilon \gamma_E}}{\Gamma(1-\epsilon)} \frac{1}{\tau_a^{\frac{2\epsilon-a}{2-a}}} \sum_{j,k} \int dx\, \left(x^{a-1}+(1-x)^{a-1}\right)^{\frac{2\epsilon-2}{2-a}} {\cal P}_{i \to jk}^{\rm med}\left(x,\frac{\tilde{p}_T\tau_a^{\frac{1}{2-a}}}{(x^{a-1}+(1-x)^{a-1})^{\frac{1}{2-a}}}\right)  \, .  
 \label{eq:medjet}
\end{align}
\end{widetext}
From \eq{eq:medjet}, we note that there are no new extra divergences introduced by the medium. Therefore, one can take the limit  $\epsilon \to 0$. As a result, at the leading order, the anomalous dimension for the angularity jet function remains the same as that of vacuum jet function. However, due to the finite contribution from the medium-induced jet function, the boundary conditions for solving the RGEs are now modified.

In the leading order of opacity expansion, the splitting function in a medium for $q$ and $g$ is given in Ref.~\cite{Ovanesyan:2011kn}. For illustrative purposes, we give below the splitting function for a light quark to a light quark and a gluon,
\begin{widetext}
\begin{eqnarray}
&&   {\cal P}^{\rm med}_{q \to qg}(x,k_{\perp})  =  C_F
  \frac{1+(1-x)^2}{x}  \!
\int \frac{d z}{\lambda_g}  
\int\! d^2{\vec q}_\perp  \frac{1}{\sigma} \frac{d\sigma}{d^2 {\vec q}_\perp} \Bigg[  \frac{\vec{B}_{\perp}}{\vec{B}_{\perp}^2}\! \cdot\!\! \left( \frac{\vec{B}_{\perp}}{\vec{B}_{\perp}^2}  -  \frac{\vec{C}_{\perp}}{\vec{C}_{\perp}^2}  \!  \right) \big( 1-\cos[(\Omega_1 -\Omega_2) z] \big)
   \nonumber \\
&&  \qquad  
      + \frac{\vec{C}_{\perp}}{\vec{C}_{\perp}^2} \cdot \left( 2 \frac{\vec{C}_{\perp}}{\vec{C}_{\perp}^2}   
-    \frac{\vec{A}_{\perp}}{\vec{A}_{\perp}^2} - \frac{\vec{B}_{\perp}}{\vec{B}_{\perp}^2}  \right)  \big(1- \cos[(\Omega_1 -\Omega_3) z] \big) + \frac{\vec{B}_{\perp}}{\vec{B}_{\perp}^2} \cdot \frac{\vec{C}_{\perp}}{\vec{C}_{\perp}^2} 
\big( 1 -  \cos[(\Omega_2 -\Omega_3) z] \big) + \frac{\vec{A}_{\perp}}{\vec{A}_{\perp}^2} \cdot \left( \frac{\vec{D}_{\perp}}{\vec{D}_{\perp}^2} - \frac{\vec{A}_{\perp}}{\vec{A}_{\perp}^2} \right) \nonumber \\  
&&
   \qquad    
 \big(1-\cos[\Omega_4 z]\big)  -\frac{\vec{A}_{\perp}}{\vec{A}_{\perp}^2} \cdot \frac{\vec{D}_{\perp}}{\vec{D}_{\perp}^2} \big(1-\cos[\Omega_5 z]\big)   
+  \frac{1}{N_c^2}  \frac{\vec{B}_{\perp}}{\vec{B}_{\perp}^2} \cdot  \left( \frac{\vec{A}_{\perp}}{\vec{A}_{\perp}^2}  -   
\frac{\vec{B}_{\perp}}{\vec{B}_{\perp}^2}      \right) \big( 1-\cos[(\Omega_1 -\Omega_2) z] \big)   \Bigg] \,, 
\label{eq:CohRadSX1} 
\end{eqnarray} 
\end{widetext}
where we have utilized the same notations as in Ref.~\cite{Ovanesyan:2011kn}. 
The quantities $\Omega_i$, $\vec{A}_{\perp}, \vec{B}_{\perp}, \vec{C}_{\perp}$ and $\vec{D}_{\perp}$ are the same as in Ref.~\cite{Ovanesyan:2011kn}. 
In \eq{eq:CohRadSX1}, the ${\it{cosine}}$ terms represent the Landau-Pomeranchuk-Migdal (LPM) effect. $\lambda_g$ is the gluon mean free path in a system of size $L$. Moreover, the normalized collision kernel ($(1/\sigma) d\sigma/d^2\vec{q}_{\perp}$) describes the elastic scattering between the jet and thermal constituents and reads as
\begin{eqnarray}
\frac{1}{\sigma} \frac{d\sigma}{d^2 {\vec q}_\perp}
=\frac{m_D^2}{\pi({q}_{\perp}^2+m_D^2)^2}\, ,
\label{infsigmael}
\end{eqnarray}
where $m_D$ is medium Debye screening mass. 

In order to obtain the final angularity distributions with the medium modifications from \eq{eq:medjet}, we further require phase space cuts on the $x$ integration. This is mainly because of an enhancement in the $x\to 0$ limit that requires opacity resummation. Progress has been made in this direction~\cite{Barata:2021wuf, Sievert:2018imd}. Nevertheless, in this study, we employ splitting functions acquired through the first-order opacity expansion. For our calculation, we follow the prescription discussed in Ref.~\cite{Sievert:2019cwq}. However, we would like to point out that the ratios of the differential distributions, defined in \eq{eq:ratio}, are not sensitive to such cuts. To obtain the resummed result for medium-modified angularity distribution, we solve \eq{eq:JsolFspace} with the total jet function containing medium-modified contributions. Moreover, as there are no large logarithms introduced in the medium-modified angularity jet function, therefore, to this order, the jet scale for RG evolution remains the same as that in the vacuum. The RG evolution equation for the total jet function ${\cal J}$ in the presence of the medium, explicitly reads as
\begin{align}
&\mu\frac{d}{d\mu}{\cal J}_{i}(y,\tilde{p}_T,\mu)
= \gamma_{{\cal J} i}(y,\tilde{p}_T,\mu) \, {\cal J}_i(y,\tilde{p}_T,\mu) \, , 
\label{eq:RGjet}
\end{align}
where $\gamma_{\mathcal{J}_i}$ is the anomalous dimension
\begin{align}
\gamma_{{\cal J} i}(y,\tilde{p}_T,\mu) &=  \Gamma_{\cal J}^{i}[\alpha_s] \Big(\frac{2}{j_J} \ln(i\, y\, e^{\gamma_E})+ \ln L_n\Big) + \gamma_{\cal J}^{i}[\alpha_s]. 
\end{align}
The boundary condition to solve Eq.~\ref{eq:RGjet} is given by the total fixed order jet function that includes both vacuum and medium-induced contributions, i.e., $\mathcal{J}_{i}^{\rm vac}+\mathcal{J}_{i}^{\rm med}$. With some simplifications, the final form of renormalized jet function acquires the form
\begin{align}
{\cal J}_i(y, \tilde{p}_T, \mu) &=  e^{K_{\gamma_J}^i+2 K_{\Gamma_J}^i} \!\! \left[(i\, y\, e^{\gamma_E})^{\frac{2\eta_{\Gamma_J}^i}{j_J}} \Big(\frac{\mu_J}{\tilde{p}_T}\Big)^{2\eta_{\Gamma_J}^i}\right]\, \nonumber \\
&\quad \times {\cal J}_i(y, \tilde{p}_T, \mu_J).  
\label{eq:JsolFspace}
\end{align}
Further details about the notation and the solution of RG equations and evolution are discussed in Appendix~\ref{app:fourier}. As mentioned earlier, we work with NLL$'$ accuracy that requires finite terms in the fixed-order distribution, while the resummed ingredients involve two-loop cusp and one-loop non-cusp anomalous dimension, discussed in Appendix~\ref{app:fourier}. For the final resummed angularity distribution we again use the factorization formula discussed in \eq{eq:AAfact} with fixed order functions replaced by resummed jet and soft functions.  

\subsection{Medium model} 
\label{sec:MediumAngularity}

In this section, we discuss the assumptions about the QGP medium under which we compute the relevant medium parameters such as the medium temperature, Debye mass, mean-free-path, and system length. We consider a simple Bjorken model for the medium, and neglect the effects of radial and transverse flow. For a fixed impact parameter $b$, medium parameters along the jet path are functions of the point of the initial hard scattering $r_0$ and the emission angle $\phi$. (There is no $\eta$ dependence as we will focus on $\eta=0$.) 
To simplify our calculation, we take the average over $\phi$ and $r_0$.  With these assumptions, for a given $b$, the average length of the encountered medium is obtained from
\begin{equation}
 L(b)= \frac{\int d^2r_0 d\phi\, L(b,r_0,\phi)\,P(b,r_0)}{\int d^2r_0 d\phi\,P(b,r_0) }   \, ,
 \label{eq:avgL}
\end{equation}
where \( L(b, \mathbf{r}_0, \phi_0) \) denotes the total path length traversed by the jet through the medium, starting from the initial hard scattering point \( \mathbf{r}_0 \) until it exits. The overlap region dominates the $r_0$ integration and $P(b,r_0)$ is the probability density of binary interaction between two nuclei which we obtain from the Glauber model. (See Ref.~\cite{Miller:2007ri} for more details.)

Since the jet travels with the speed of light, one can replace the time with the distance traversed by the jet, and hence when the jet has covered a distance $L$ in the medium, the temperature is defined as
\begin{equation}
T(b,r_0,\phi)=T_0(b)\bigg(\frac{L(b,r_0,\phi)}{\tau_0}\bigg)^{\frac{1}{3}} \, ,
\end{equation}
where $\tau_0$ is same as thermalization time which we take as $0.6$ fm~\cite{Chang:2015hqa}. We ignore the short distance traveled in the first $0.6$ fm. It is to be noted that in the above equation, we have kept the angular dependence of $T_0$ and initial position. For our purpose, we assume that the initial temperature is independent of $\phi$ . For initial temperature ($T_0$), we follow the prescription provided in Ref.~\cite{Sharma:2023dhj}. Thus, with all these ingredients, the average temperature is given by
\begin{equation}
 T(b)=\frac{\int d^2r_0 d\phi  T(b,r_0,\phi){P}(b,r_0)}{\int d^2r_0d\phi {P}(b,r_0)}\, . 
\label{eq:temp}
\end{equation}
From the average temperature, we estimate the average Debye mass ($m_D=\sqrt{1+N_f/6}\,g_s T$, where $N_f$ is number of flavors) for each impact parameter $b$.
The mean free path of a gluon is evaluated as~\cite{Ke:2022gkq}
\begin{equation}
\lambda_g^{-1} = \frac{9\zeta(3)}{32\pi^3} \frac{g_s^4\,T^3}{m_D^2}\left(16+4 N_f\right)\,.
\label{eq:mpath}
\end{equation}
For quarks, $\lambda_q^{-1}=(C_F/C_A) \lambda_g^{-1}$.
Finally, we integrate $b$ over the ranges that correspond to the relevant centrality bins.

In Table~\ref{tab:medmodel} we tabulate medium parameters obtained using \eqns{eq:avgL}{eq:temp}{eq:mpath}. To obtain the Debye mass, we take $g_s=2$. We note from Table~\ref{tab:medmodel} that the central value for the average temperature for the 10-30\% centrality bin is slightly larger than the 0-10\% centrality bin, even though the initial temperature is larger for the more central bin. This is because the reduced initial temperature for the more peripheral bin happens to be compensated by the fact that jets spend a smaller time in the cooling medium.
\begin{table}[t]
\centering
\renewcommand{\arraystretch}{1.5}
 \begin{tabular}{|c|c|c|} 
 \hline
 parameter & $0-10\%$ & $10-30\%$ \\ [0.4ex] 
 \hline
 $b$(fm) & 3.34 & 7.01  \\
 $L$(fm) & 4.96$\pm$ 2.44 & 3.56$\pm$ 1.98  \\
 $T_0$(MeV) & 456 & 437  \\
 $T$(MeV) & 248$\pm$ 128 & 289$\pm$ 84  \\
 $m_D$(MeV) & 0.607$\pm$ 0.314 & 0.708$\pm$ 0.206  \\
 $\lambda_g^{-1}$ (fm$^{-1}$) & 1.009$\pm$ 0.52 & 1.17$\pm$ 0.34  \\ [1ex] 
 \hline
\end{tabular}
\caption{Medium parameters for PbPb collisions obtained from Glauber model for
$0-10\%$ and $10-30\%$ centrality bins. The initial temperature ($T_0$) is
obtained following the method discussed in Ref.~\cite{Sharma:2023dhj}.}
\label{tab:medmodel}
\end{table}
This feature is model dependent and a Bjorken-like expansion model is a bit simplistic to capture correctly these dynamics as would be the case typically in a more sophisticated hydrodynamical model. We would, however, like to point that the temperatures between the two bins are still compatible within the given error bands. For the plots presented in this paper, we will only use the central values of the medium parameters and show the uncertainties due to variation of the renormalization scale $\mu$ only. 

With medium parameters in hand, we can use
\eqs{eq:medjet}{eq:CohRadSX1} to calculate the jet function for a
given $p_T$. However, to evaluate the final angularity distribution (Eq.~\ref{eq:AAfact}), we
need to account for two additional effects. First, the jet $p_T$ is modified by interactions with the medium partons (energy loss), causing a migration effect in the observed $p_T$ of the jet. 
Second, the parton distribution functions
in AA collisions are different from the distribution functions
in $\rm pp$ collisions due to initial state effects. These processes affect the overall jet $p_T$ distribution and therefore its moments. We describe these effects in the next section.

\subsection{Jet $p_T$ spectrum~\label{sec:elossdetails}}

Jet energy loss in the medium is larger than in vacuum due to the interactions that jet partons undergo with the thermal constituents, leading to additional out-of-cone radiation. This implies that a jet measured in a given $p_T$ range can have contributions from various higher $p_T$ initiated jets that have lost a sufficient amount of energy~\cite{Brewer:2020tbb,Brewer:2020chg}. For instance, a high $p_T$ jet can lose a substantial fraction of its energy and therefore can migrate to the observed bin. Moreover, a jet initiated by a parton with slightly larger $p_T$, that lost only a small fraction of the energy, can end up within the same observed $p_T$ bin. This effect needs to be included while comparing jet substructure observables in heavy-ion collisions with $\rm pp$ collisions~\cite{Rajagopal:2016uip}. 
 
The average energy loss fraction via a single out-of-cone emission, for gluon- ($\bar{\varepsilon}_g$) and quark- ($\bar{\varepsilon}_q$) initiated jet in the medium, can be computed~\cite{Chien:2015hda} from the medium-induced splitting
functions ${\cal{P}}^{\rm{med}}$ as follows:
\begin{align}
\bar{\varepsilon}_g&\!=\!2 \pi \left[\int_0^{\frac{1}{2}}dx \, x +\int_{\frac{1}{2}}^1 dx \, (1-x)\right] \!\int_{2 p_T x(1-x)\tan{\frac{R}{2}}}^{2 p_T x(1-x)\tan{\frac{R_0}{2}}}\!\!\!\!  dk_{\perp}\nn \\
&\quad \times k_{\perp}\!\left[\mathcal{P}^{\rm med}_{g \to gg}(x,k_{\perp})+2 N_f \mathcal{P}^{\rm med}_{ g \to q\bar{q}}(x,k_{\perp}) \right] , \label{eq:geloss}  \\~\nn \\ 
\bar{\varepsilon}_q&\!=\! 2 \pi \left[\int_0^{\frac{1}{2}}dx \, x+\int_{\frac{1}{2}}^1 dx \, (1-x)\right] \int_{2 p_T x(1-x)\tan{\frac{R}{2}}}^{2 p_T x(1-x)\tan{\frac{R_0}{2}}}\!\!\!\!  dk_{\perp}\nn \\ 
&\quad \times k_{\perp}\left[\mathcal{P}^{\rm med}_{q \to qg}(x,k_{\perp})+ \mathcal{P}^{\rm med}_{q \to gq}(x,k_{\perp}) \right]   \, . \label{qeloss}
\end{align}
The upper limit on $k_{\perp}$ in Eqs.~\ref{eq:geloss},~\ref{qeloss} is determined by $R_0 \sim {\cal O}(1)$.

\begin{figure}[t]
\centering
\hspace{-0.35cm}\includegraphics[scale=0.85]{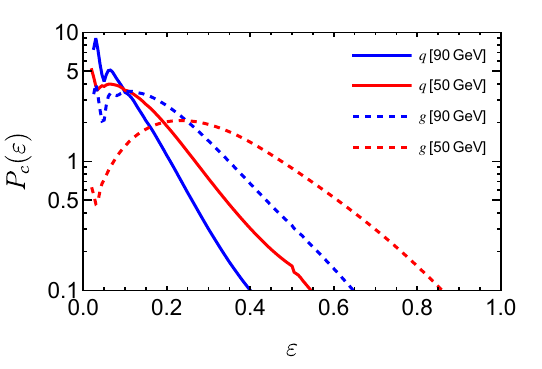}
\caption{Distributions of fractional energy loss of quark ($q$) and gluon ($g$) initiated jets for initial energy $50$\;GeV and $90$\;GeV. The distributions are obtained by modeling medium-induced radiation through medium-induced splitting functions and accounting for multi-gluon emissions via a Poisson ansatz. The mean value of $\varepsilon$ gives the average energy loss values as obtained via \eqs{eq:geloss}{qeloss}, and the variance represents fluctuations in the energy loss. We note that gluon-initiated jets experience larger fluctuations due to a longer tail in the distribution which is especially pronounced at small $p_T$ values. } 
\label{fig:Eloss_5090}
\end{figure}

\begin{figure}[t]
\centering
\hspace{-0.35cm}\includegraphics[scale=0.5]{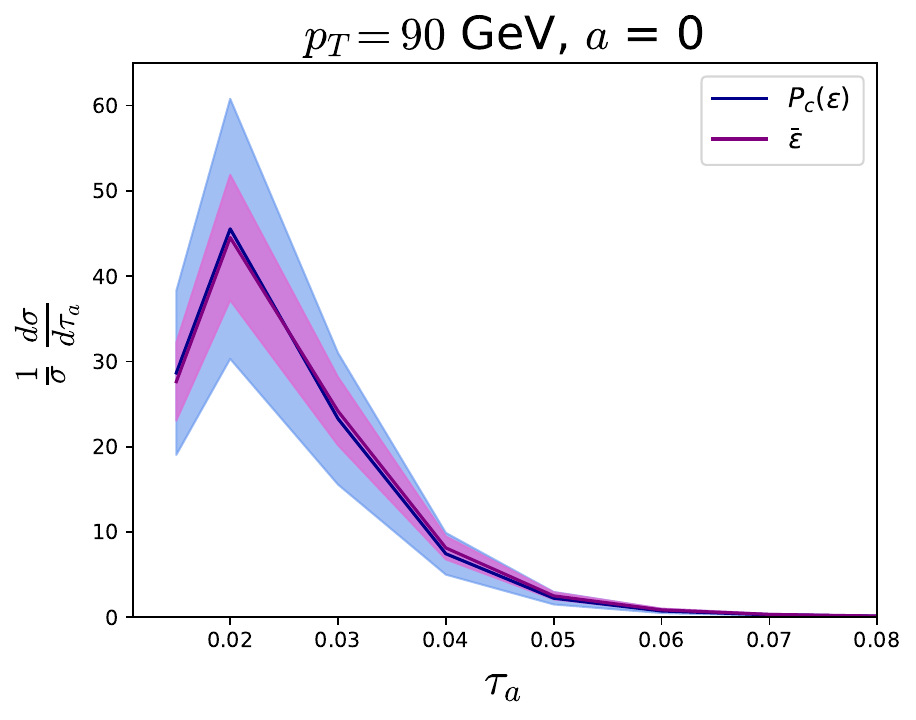}
\caption{Normalized differential distributions for $a=0$ for $p_T=90\, \rm{GeV}$, for jets with $R=0.4$ and for the $0-10\%$ centrality bin in PbPb collisions. Normalization is same as the one used in Fig.~\ref{fig:pp_compare}. The magenta error band corresponds to variation in $\mu$ from $p_T/2$ to $2p_T$, with the central line corresponding to $\mu=p_T$, taking $\varepsilon$ to be $\bar{\varepsilon}$. The blue band incorporates the full Poisson distribution, accounting for the fluctuations in the gluon number density. } 
\label{fig:avg_vs_full}
\end{figure}

As a first approximation, we include this energy loss by setting $P_c(\varepsilon)$ equal to $\delta(\varepsilon-\bar{\varepsilon})$ in \eq{eq:AAfact} (i.e. use the average value of the energy loss fraction). This provides an estimate of the energy loss due to the first moment of the radiative spectrum. 
From \eq{qeloss}, we obtain around $10-15\%$ ($20-30\%$) energy loss for quark-initiated (gluon-initiated) jets with a $p_T$ in the range $40-60$ GeV and $R=0.4$ for $0-10\%$ central collisions. 
The results for $p_T\in[80,100]$ GeV ($p_T\in[40,60]$ GeV)
 by employing the average energy loss is shown in the left (right) panel of Fig.~\ref{fig:compare_pT}, in Section~\ref{sec:results}.
\begin{figure*}[t]
\hspace{-0.5cm}\includegraphics[scale=0.7]{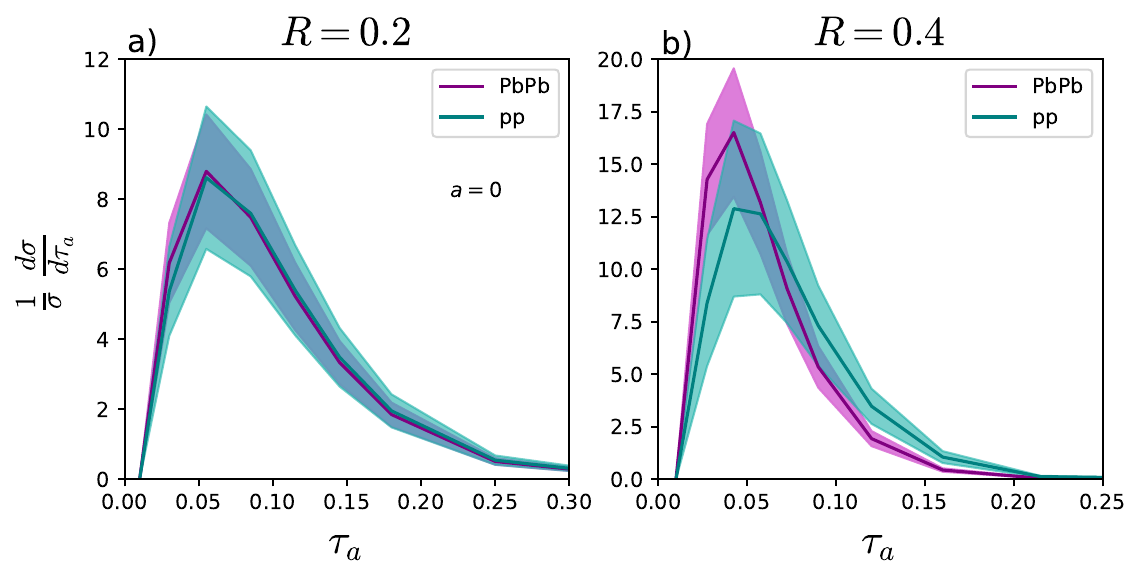}
\caption{Differential distributions in PbPb (0-10\% centrality) and pp  for $a=0$, jet parameters $40<p_T<60\, \rm{GeV}$ and (a) $R=0.2$, (b) $R=0.4$. The turquoise region represents the angularity distribution in vacuum, and the magenta region is the medium-modified distribution. The theoretical error band corresponds to variation in $\mu$ from $p_T/2$ to $2p_T$, with the central line corresponding to $\mu=p_T$. The non-perturbative parameter $\Omega_0$ is $0.85(0.8)\, \rm{GeV}$ for $R=0.2(0.4)$. See the text below \eq{eq:Fk} for more details. Normalization is same as the one in Fig.~\ref{fig:pp_compare}.}
\label{fig:compare_R}
\end{figure*}

This simple picture is complicated by two important effects. First, even for a fixed trajectory of a jet in the medium, the energy carried outside the cone by a parton is not fixed. The splitting functions give the (unnormalized)
probability distribution of the energy loss fraction, and the variance in the distribution about the mean can be substantial. Furthermore, the number of partons emitted out-of-cone is not fixed. A traversing energetic parton can emit
multiple partons as it moves through the medium (multiple splitting of gluons in $q\bar{q}$ is small and can be neglected). These effects can be modeled by assuming that the gluon emissions are independent, and by writing the multi-gluon emission distribution as a Poisson distribution~\cite{Gyulassy:2001nm}. We show the distribution of fractional energy loss, $P_c(\varepsilon)$, for two representative $p_T$ values in Fig.~\ref{fig:Eloss_5090}.  

In Fig.~\ref{fig:avg_vs_full}, to consider the effect of accounting for the full energy loss distribution, we compute the final angularity distribution by using $P_c(\varepsilon)$ in Eq.~\ref{eq:AAfact} and compare the result with what we obtain if we use just the mean energy loss. From the figure, we see that the central values of the angularity distribution calculated with the full $P_c(\varepsilon)$ (blue) is close to the result obtained by using the first moment of $\bar{\varepsilon}$ (magenta) while the uncertainty bands are wider, as expected.  The uncertainty bands in the blue are obtained by varying the scale $\mu$ from $p_T/2$ to $2p_T$ and convoluting the distribution with $P_c(\varepsilon)$. The calculation of the angularities using the full distribution is computationally quite time-consuming. Therefore, for the rest of the plots in this paper (Figs.~\ref{fig:compare_R}, ~\ref{fig:compare_pT},~\ref{fig:compare_a},~\ref{fig:compare_cent}), we will use the average energy loss 
for the jet in a given $p_T$ bin. We expect that this will give a good estimate of normalized distribution for the central value, but we underestimate the error band in the angularity distributions.

The second source for fluctuations in the energy loss is that for a given centrality bin, one needs to average over the impact parameters, the jet production point, and the direction relative to the medium, all of which will affect the temperature and path lengths encountered by the jet, see
\eq{eq:CohRadSX1}. Thus the energy loss distribution needs to be computed on a jet-by-jet basis. Additionally, in fluctuating hydrodynamical backgrounds, averages over
the background must be performed. A detailed investigation of all these effects is interesting and will be explored in future.

Finally, the initial state effects can be accommodated by modifications of the parton distribution functions $\tilde{f}(x,\mu)$ from their vacuum values.
The key process is the initial state interactions of the colored initial hard parton as it traverses through the second nucleus, prior to the hard interaction with the
opposite moving parton ~\cite{Vitev:2007ve}. This is otherwise known as the cold nuclear matter energy loss. This energy loss is process-dependent, and it has been argued that it is not fully captured in the nuclear PDFs~\cite{Vitev:2006bi}. We use the model for CNM energy loss given in Ref.~\cite{Vitev:2007ve}, which has been previously used for the phenomenology of jets in heavy-ion collisions. 

With all the ingredients in hand, we now calculate the effect of medium modifications on the differential angularity distributions for various jet cone size, jet $p_T$ and centrality. 

\subsection{Discussion of medium-modified angularity distributions}
\label{sec:results}

\begin{figure*}[t]
\hspace{-0.5cm}\includegraphics[scale=0.7]{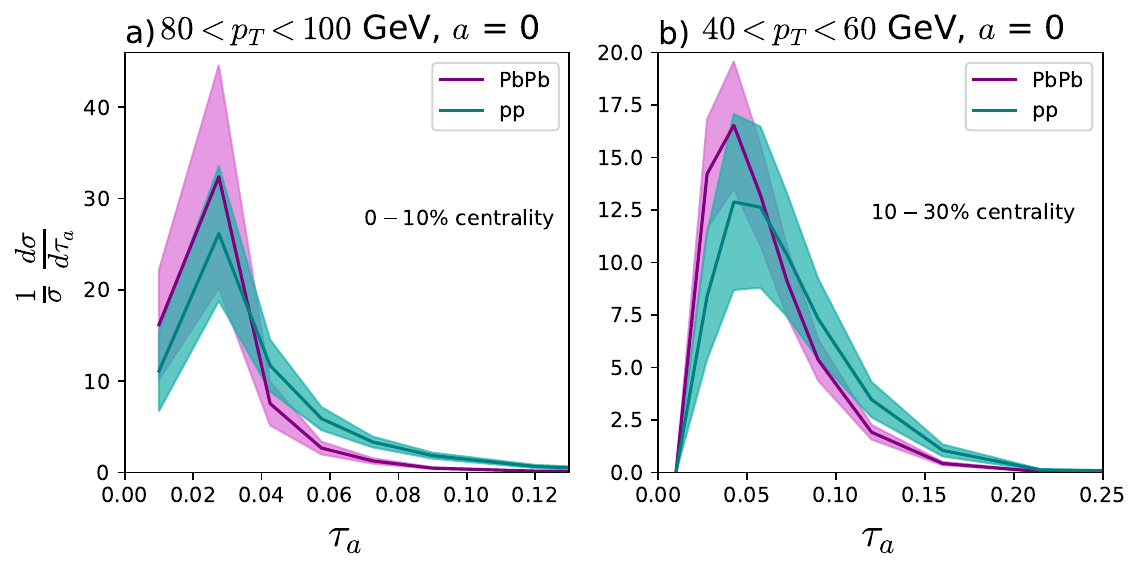}
\caption{Differential distributions for $a=0$ with (a) $80<p_T<100\, \rm{GeV}$, $0-10$\% centrality and (b) $40<p_T<60\, \rm{GeV}$, $10-30$\% centrality, for a jet with $R=0.4$. The non-perturbative parameter is taken as; $\Omega_0=0.35\, \rm{GeV}$ for the left plot and $\Omega_0=0.8\, \rm{GeV}$ for the right one.  Normalization is same as the one in Fig.~\ref{fig:pp_compare}.}
\label{fig:compare_pT}
\end{figure*}

There are two effects that control the distribution of medium-modified angularities. First, the energy loss experienced by a jet while traversing the medium, due to out-of-cone radiation, resulting in a $p_T$ shift of the jet initiating parton.  This leads to an overall shift in the distributions for all exponents. Second, the medium-induced splittings can generate local changes in the distribution of the observable leading to a further redistribution of the spectrum.  The combined effect of these two depends on the jet parameters for a given set of medium scales as discussed below. 

In Figure~\ref{fig:compare_R}, we show differential distributions normalized to unity for angularity exponent $a=0$ and two different values of jet radius i.e., $R=0.2$ and $R=0.4$, for a jet with momentum $40<p_T<60\, \rm{GeV}$ and centrality $0-10$\%.   
As may be seen from Figure~\ref{fig:compare_R} (a), a narrower jet ($R=0.2$) is less modified in the whole range of $\tau_a$ in contrast to $R=0.4$. 
To the leading order, for the medium, the soft function is $R$ dependent while the jet function is less sensitive to the cone size. Therefore, smaller $R$ jets are 
less modified in our analytic setup.   
On the other hand, for $R=0.4$, there is an apparent shift of the distribution towards small $\tau_a$ region arising from different competing effects. Additionally, in the larger $\tau_a$ region the medium-modified distribution is steeper compared to the vacuum one.      

Next, we analyze the effect of medium modification on jets (i) initiated by a parton with a different $p_T$, and (ii) created in off-central collisions.  To this end, in Figure~\ref{fig:compare_pT}, we present differential distributions for jets with $R=0.4$, (a) $80<p_T<100\, \rm{GeV}$ with centrality $0-10$\% and (b) $40<p_T<60\, \rm{GeV}$ with centrality $10-30$\%, keeping the differential distribution of Figure~\ref{fig:compare_R} (b) as the baseline. For a high $p_T$ jet, the medium modifications are less apparent. However, the distributions in the medium are still enhanced in the small $\tau_a$ region, and a steeper fall is seen for larger values of $\tau_a$. 

\begin{widetext}
\begin{figure*}[t]
\hspace{-0.5cm}\includegraphics[scale=0.7]{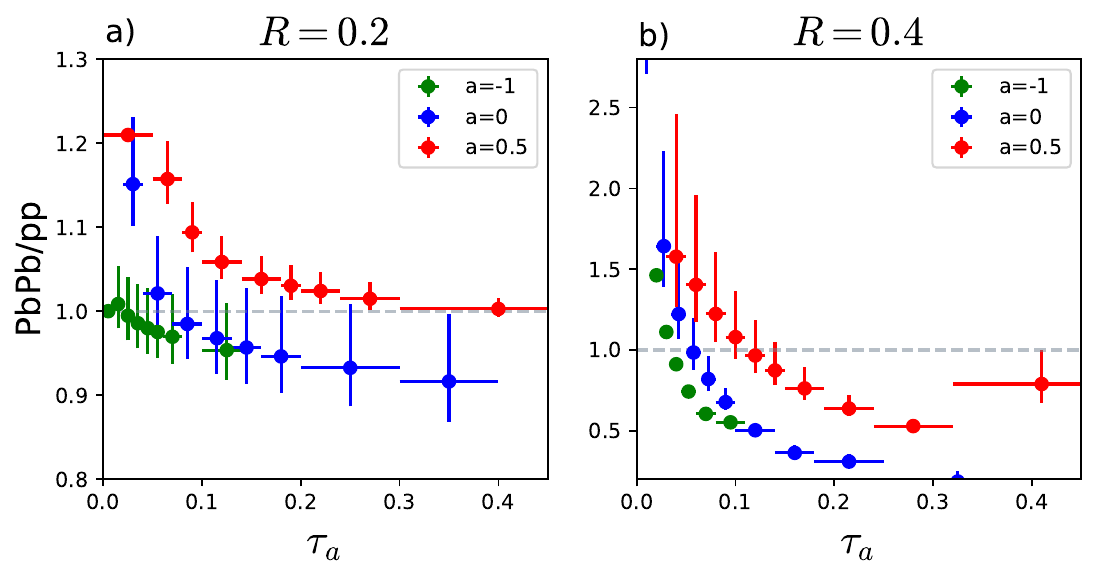}
\caption{Ratio of differential distributions in  PbPb and pp for various angularity exponents $a=-1,0,0.5$;  (a) for $R=0.2$ and (b) for $R=0.4$. For both the figures the jet $p_T$ is in the range $40<p_T<60\, {\rm GeV}$ and centrality is $0-10$\%.  As can be seen, medium effects are stronger in jets with a larger radius for all the angularity exponents considered here.} 
\label{fig:compare_a}
\end{figure*}
\end{widetext}

\begin{figure}[h]
\includegraphics[scale=0.5]{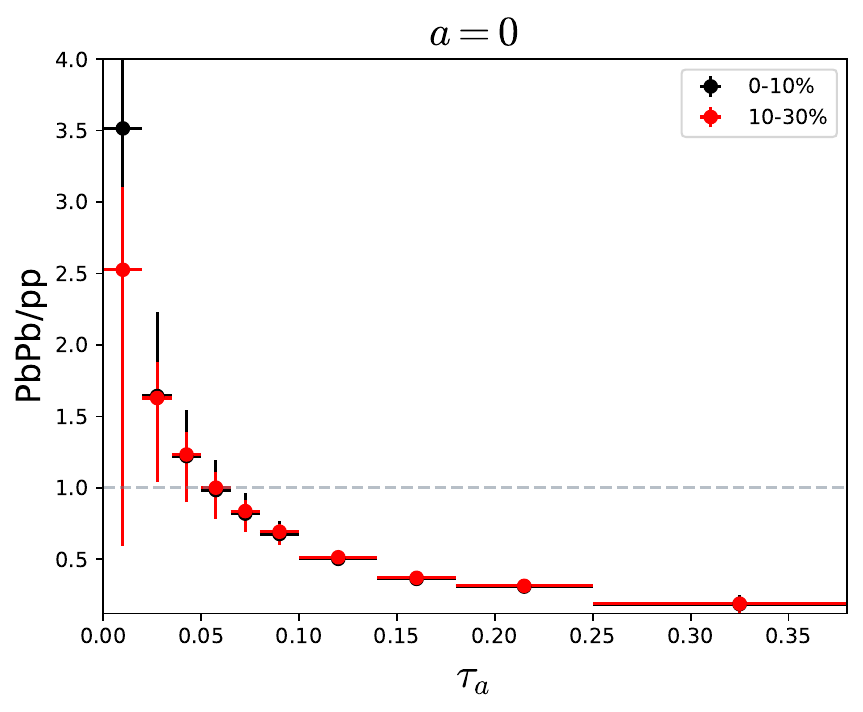}
\caption{Comparison of ratios of differential angularity cross-section for PbPb to pp for two centrality bins $0-10$\% (black) and $10-30$\% (red). Jets are clustered with $R=0.4$ and transverse momentum in the range $40<p_T<60\, {\rm GeV}$. The result is shown for the angularity exponent $a=0$ which is also related to jet mass.} 
\label{fig:compare_cent}
\end{figure}

To see the effect of off-central collisions, from Figure~\ref{fig:compare_pT} (b), we find 
that the differential distribution for the 10-30\% centrality bin is quite similar to the central case, and we see only a slight modification of the medium-modified distribution compared to the central case. To understand it better, we present the ratios of medium-modified and vacuum differential angularity distributions as a function of $\tau_a$ for the exponent $a=0$ in Figure~\ref{fig:compare_cent}. Interestingly, we see very little difference between 0-10\% and 10-30\% bins, except for the small $\tau_a$ region which shows less enhancement for the latter case. The fact that the difference in all other $\tau_a$ bins is not so significant can be traced back to the parameters found in Table~\ref{tab:medmodel} for the average temperature. It is important to note that this feature is, however, reliant on the use of a Bjorken-like model for the hydrodynamic evolution of the medium. However, the feature that in the off-central case, jets would be less modified (as visible in the small $\tau_a$ region) than the more central case, is robust and does not rely on the specific details of a model. 
Even though at first sight, this seems not so obvious, a similar feature is observed when looking at another jet property, the nuclear modification factor, R$_{AA}$ for these two centrality bins. This is partially seen in the available experimental data from CMS and ATLAS~\cite{ATLAS:2014ipv,ATLAS:2018gwx, CMS:2016uxf} for $p_T$ not much greater than 100 GeV, towards the lower end of the available $p_T$ bins explored in these experiments. In this region, the R$_{AA}$ of the 0-10\% central bin seems to be not very different from the next centrality bin. A more thorough understanding of R$_{AA}$ for $p_T$ values we consider in this paper requires extending the experimental data to small $p_{T}$ jets in the same centrality bins. 

Finally, to qualitatively understand the impact of varying the exponent `$a$', we show the ratio of angularity distributions in PbPb and pp collisions in Figure~\ref{fig:compare_a} for a jet with $40<p_T<60\, {\rm GeV}$ and for two different jet radius parameters $R=0.2$ and $R=0.4$.  We present these results for three different angularity exponents, namely, $a=-1,\, 0\,$ and $0.5$.  The corresponding quantity shown in the figure is defined as
\begin{equation}
\frac{\text{PbPb}}{\text{pp}}=\frac{\left(\frac{1}{\sigma}\frac{d\sigma}{d\tau_a}\right)_{\text{AA}}}{\left(\frac{1}{\sigma}\frac{d\sigma}{d\tau_a}\right)_{\text{pp}}}, 
\label{eq:ratio}
\end{equation}
where each of the distributions are normalized to unity, respectively. 

From Figure~\ref{fig:compare_a}, we see that a narrower jet with cone size $R=0.2$ is much less modified in contrast to a jet with radius $R=0.4$. Furthermore, the smaller angularity exponents, which probe the collimated structure of the jet, are less sensitive to medium dynamics. In particular, the distribution for $a=0.5$ shows the largest modification due to a higher sensitivity to the spread of a jet, for both the jet radius parameters. In fact, the jet substructure modifications due to the medium become more pronounced for all the `$a$' values for $R=0.4$. A combined scan of different angularity exponents and different jet radius parameters over various centrality bins could therefore be a useful probe towards a better understanding of medium-induced effects. 


\section{Summary}
\label{sec:discussion}

In this paper, we performed a first study of medium modifications on ungroomed jet angularities, restricting to the range of $a<1$ exponents. Angularities are interesting observables in the sense that one can vary the sensitivity of the observable to collinear radiation in the jet by varying the exponent `$a$'. We present results for angularity exponents $a=-1,\, 0$ and $ 0.5$, although the framework holds for any value of `$a$' less than 1. To describe jet-medium interactions, we utilize the SCET$_{\rm G}$ formalism that incorporates these interactions via background Glauber fields generated from the color charges in the QGP. Utilizing the collinear matrix elements and the medium-induced splitting functions, we calculate the medium-modified angularity jet function to NLL$^\prime$ accuracy. We use nCTEQ15 nuclear PDFs for Pb nuclei and incorporate the initial state energy loss before the hard interaction using the formalism discussed in Ref.~\cite{Vitev:2007ve}. To capture medium effects, we estimate medium parameters within the Glauber model and obtain the differential angularity spectrum.

In order to explore the medium modifications on the jet substructure, we present results for various jet parameters, angularity exponents, as well as for an off-central event. By looking at two different values of jet radius and two different centrality bins, we find that the effects of the medium are more pronounced for $a=0.5$ and for a wider jet cone. Although we see little-to-no differences between the two centrality bins in our study, this effect is a result of the simple Bjorken-like expansion of the medium considered here. We also note from Table~\ref{tab:medmodel} that the fluctuations in the medium parameters within each centrality bin are quite large. Therefore, a proper analysis will require taking these into account. Further, one can use more sophisticated models of the background hydrodynamic evolution to characterize the medium. These improvements we leave for future work.

\acknowledgments

We acknowledge the support of the Department of Atomic
Energy, Government of India, under Project Identification No. RTI 4002. We thank Ivan Vitev, Felix Ringer and Sidharth Kumar Prasad for their useful comments on the manuscript. We also thank Jean-Philippe Guillet for providing Fortran code for the hard function. We acknowledge the
computational facility provided by the Department of Theoretical Physics at TIFR.


\appendix

\section{Plus Distributions}
\label{app:PlusDistributions}
For a generic function $g(x)$ less singular than $1/x^2$ as $x \to 0$, the plus distribution with a boundary at $x=x_0$ is defined as~\cite{Ligeti:2008ac}
\begin{equation}
\label{appeq:PlusDefine}
\left[\theta(x)\, g(x)\right]_{+}^{x=x_0} = \lim_{\e \rightarrow 0}\Big[\theta(x-\e)\, g(x) + \delta(x-\e)\!\! \int_{x_0}^x\! \df x'\, g(x')\Big],
\end{equation}
where the boundary condition satisfies
\begin{equation}
\label{eq:BoundaryConditionDefine}
\int_{0}^{x_0}\! \df x\, \left[\theta(x)\, g(x)\right]_{+}^{x=x_0} = 0\, .
\end{equation}
The boundary condition of the plus distribution can be related to a plus distribution with a different boundary through the relation
\begin{equation}
\label{appeq:BoundaryChangeRelation}
[\theta(x)\, g(x)]_{+}^{x=x_0} = [\theta(x)\, g(x)]_{+}^{x=1} + \delta(x)\!\! \int_{x_0}^1\!\! \df x'\, g(x') .
\end{equation}
In particular, one can rewrite this relation for the distributions where one has a boundary at infinity to express in terms of the one with a boundary at $x=+1$. 
\begin{equation}
\label{appeq:PlusExpand}
\left[\!\frac{\theta(x)}{x^{1+\alpha}}\!\right]_{+}^{\infty} \!\!\!\!= -\frac{1}{\alpha}\delta(x)+\!\left[\!\frac{\theta(x)}{x}\!\right]_{+}\!\!\!\!-\alpha \!\left[\!\frac{\theta(x) \ln x}{x}\!\right]_{+}\!\!\! + {\cal O}(\alpha^2)\, .
\end{equation}
Another useful identity  that we require is the {\emph{rescaling}} identity, i.e., any constant parameter $\zeta$ (with $\zeta > 0$) can be scaled out of the distribution as 
\begin{align}
\label{appeq:PlusFunction}
\frac{1}{\zeta}\left[\frac{\theta(x)}{x/\zeta}\right]_{+} &= \left[\frac{\theta(x)}{x}\right]_{+}\!\!\! - \ln \zeta\, \delta(x)\, , \\
\frac{1}{\zeta}\left[\frac{\theta(x)\ln(x/\zeta)}{x/\zeta}\right]_{+} &= \left[\frac{\theta(x) \ln x}{x}\right]_{+}\!\!\!\!\! - \ln\zeta \left[\frac{\theta(x)}{x}\right]_{+}\!\!\!\!\! \nonumber \\
&\quad + \frac{1}{2} \ln^2\! \zeta\,\, \delta(x)\, .
\end{align}
\vspace{0.5em}

\section{Perturbative results}\label{app:pert}
In this section, we collect some necessary ingredients for our perturbative computations that were left out of the main text.

\subsection{Fixed-order computation of the vacuum jet function}\label{app:Jet}
For a quark-initiated jet, the angularity jet function in vacuum reads as
\begin{align}
&{\cal J}_q^{\rm vac}(\tau_a,p_T,\mu) 
= \frac{\bar{\alpha}_s}{\vert2-a\vert} \frac{e^{\epsilon \gamma_E}}{\Gamma(1-\epsilon)} \left(\frac{\mu^2}{p_T^2}\right)^{\!\!\epsilon} \frac{1}{\tau_a^{1+\frac{2\epsilon}{2-a}}} \nonumber \\
& \quad \times \int dx\, {\cal P}_{q \to gq}(x) \left(x^{a-1}+(1-x)^{a-1}\right)^{\frac{2\epsilon}{2-a}} . 
\end{align}
Simplifying this by factoring out the term $(1-x)^{-2\epsilon\frac{1-a}{2-a}}$ to regulate the divergence at $x\to 1$ in the splitting function, we obtain the form
\begin{align}
&{\cal J}_q^{\rm vac}(\tau_a,p_T,\mu)=\frac{\bar{\alpha}_s C_F}{\vert2-a\vert} \frac{e^{\epsilon \gamma_E}}{\Gamma(1-\epsilon)} \left(\frac{\mu^2}{p_T^2}\right)^{\!\!\epsilon} \frac{1}{\tau_a^{1+\frac{2\epsilon}{2-a}}} \nonumber \\
&\qquad \int\!\! dx\, \Big(1+\Big(\frac{x}{1-x}\Big)^{a-1}\Big)^{\!\!\frac{2\epsilon}{2-a}} \times \Big(\frac{1+x^2}{(1-x)^{1+2\epsilon\frac{1-a}{2-a}}}\nonumber \\
&\qquad - \epsilon (1-x)^{1-2\epsilon\frac{1-a}{2-a}}\Big) \, .
\end{align}
Now, performing the $x$ integral by first expanding in $\epsilon$ and keeping only terms up to ${\cal O}(\epsilon)$ in the above expression, we obtain the final form of NLO quark jet function that reads as
\begin{widetext}
\begin{align}
{\cal J}_q^{\rm vac}(\tau_a,p_T,\mu) &=\frac{\alpha_s(\mu)\, C_F}{\pi\vert2-a\vert} \bigg(\delta(\tau_a) \Big(-\frac{2-a}{2\epsilon} -\frac{2-a}{2} L_n -\frac{\epsilon}{4}(2-a)L_n^2 + \frac{\pi^2}{24} \epsilon (2-1) \Big) + \left[\frac{1}{\tau_a}\right]_{+}\!\!\!\! +\epsilon L_n\left[\frac{1}{\tau_a}\right]_{+} \nonumber \\
& \quad  -\frac{2\epsilon}{2-a} \left[\frac{\ln\tau_a}{\tau_a}\right]_{+}\bigg) \times \bigg\{-\frac{2-a}{1-a}\frac{1}{\epsilon} -\frac{3}{2} -\epsilon\Big(\frac{1}{2}+3\frac{2(1-a)}{2-a}\Big) +\frac{2\pi^2 \epsilon}{3} \frac{1-a}{2-a} + \frac{2\epsilon}{2-a} \nonumber \\
&\quad \times \int_0^1 dx\, \frac{1+(1-x)^2}{x} \ln[x^{1-a}+(1-x)^{1-a}] \bigg\} \, ,
\end{align}
\end{widetext}
which agrees with the one obtained in Ref.~\cite{Kang:2018qra}. For the gluon-initiated jet, we follow similar steps by taking the contributions from ${\cal P}_{gg}$ and ${\cal P}_{qg}$.

\subsection{Fourier space solution of jet and soft functions}\label{app:fourier}
For the jet and the soft functions, the RG equations have the form of a convolution in the measurement variable $\tau_a$, in the physical space. In order to solve these coupled RG equations, we define a Fourier transform with respect to the observable such that the RG equations become multiplicative in the Fourier space. The Fourier transform of a generic function $F$ depending on a variable $x$ is defined as 
\begin{equation}
    F(y) = \int_0^{\infty} dx\, e^{-i x y} F(x)\, .
\end{equation}
The RG equation for the soft function, in the Fourier space, then read as
\begin{align}
\mu\frac{d}{d\mu}\tilde{\cal S}_{i}(y,p_T,\mu) 
= \gamma_{{\cal S} i}(y,p_T,\mu) \, \tilde{\cal S}_i(y,p_T,\mu)\, ,
\label{eq:RGFspace}
\end{align}
where the anomalous dimension is given as
\begin{align}
    \gamma_{{\cal S} i}(y,p_T,\mu) &= -2\, \Gamma_S^i[\alpha_s] \Big(\!\ln L_s+\frac{1}{j_S} \ln(i\, y\, e^{\gamma_E})\Big).
    \label{eq:Fanomdim}
\end{align}
Here $\Gamma$ and $\gamma$ are the cusp and non-cusp anomalous dimensions. These exhibit an order-by-order
expansion in the strong coupling. To the desired NLL$^{\prime}$ accuracy, we require up to two-loop coefficient of the cusp piece, the one-loop non-cusp anomalous dimension, and two-loops running of the beta function. The expansions of the anomalous dimension pieces as well as the beta function in powers of $\alpha_s$ are written as
\begin{align}
\Gamma[\alpha_s] =& \sum_{n=0}^{\infty}\Gamma^n \Big(\frac{\alpha_s}{4\pi}\Big)^{n+1}  \, , 
\gamma[\alpha_s] =& \sum_{n=0}^{\infty}\gamma^n \Big(\frac{\alpha_s}{4\pi}\Big)^{n+1} \, , \,\,\nonumber\\  
\beta[\alpha_s] =& -2\alpha_s \sum_{n=0}^{\infty} \beta^n \Big(\frac{\alpha_s}{4\pi}\Big)^{n+1} \, .
\end{align}
Up to two loops, the $\beta$-function coefficients that we require are given as (in $\overline{\rm MS}$ scheme)
\begin{align}
    & \beta_0 = \frac{11}{3}C_A-\frac{4}{3}T_R\,N_f \, ,\\
    & \beta_1 = \frac{34}{3}C_A^2-\left(\frac{20}{3}C_A+4\, C_F\right)T_R\,N_f\, .
\end{align}
Having the anomalous dimensions as given in \eq{eq:Fanomdim}, the solution of the RG equation for the soft function, in the Fourier space, reads as
\begin{align}
\tilde{\cal S}_i(y, p_T, R, \mu) &= e^{\!-2 K_{\Gamma_S}^i} \!\! \left[(i\, y\, e^{\gamma_E})^{-\frac{2\eta_{\Gamma_S}^i}{j_S}} \Big(\frac{\mu_S R^{1-a}}{p_T}\Big)^{\!-2\eta_{\Gamma_S}^i}\right]\, \nonumber \\
&\quad \times \tilde{\cal S}_i(y, p_T, R, \mu_S). \label{eq:SsolFspace}
\end{align}

Note that both jet and soft functions are evolved from their corresponding natural scales $\mu_{J,S}$ to an arbitrary renormalization scale $\mu$ which we choose to be the hard matching scale $\mu_{\cal H} \sim p_T R$.
$K_{\Gamma}$, $k_{\gamma}$ and $\eta_{\Gamma}$ are the evolution kernels that depend on the characteristic scale of the relevant dynamics described and the free renormalization scale. For brevity, we have suppressed this scale dependence in \eqs{eq:JsolFspace}{eq:SsolFspace}.
These evolution kernels to all-orders are defined as

\begin{align} 
K_\Gamma^i(\mu_0, \mu)
=&\int_{\alpha_s(\mu_0)}^{\alpha_s(\mu)}\!\frac{\df\alpha_s}{\beta(\alpha_s)}\,
\Gamma^i[\alpha_s] \int_{\alpha_s(\mu_0)}^{\alpha_s} \frac{\df \alpha_s'}{\beta(\alpha_s')}
\,,\nonumber\\
\eta_\Gamma^i(\mu_0, \mu)
=& \int_{\alpha_s(\mu_0)}^{\alpha_s(\mu)}\!\frac{\df\alpha_s}{\beta(\alpha_s)}\, \Gamma^i[\alpha_s]
\,,\nonumber \\
K_\gamma(\mu_0, \mu)
=& \int_{\alpha_s(\mu_0)}^{\alpha_s(\mu)}\!\frac{\df\alpha_s}{\beta(\alpha_s)}\, \gamma(\alpha_s)
\, ,
\label{eq:Keta_def}
\end{align}
and their explicit expressions up to NLL$^{\prime}$ are
\begin{align} \label{eq:Keta}
K_\Gamma(\mu_0, \mu) &= -\frac{\Gamma_0}{4\beta_0^2}\,
\biggl\{ \frac{4\pi}{\alpha_s(\mu_0)}\, \Bigl(1 - \frac{1}{r} - \ln r\Bigr)\nonumber\\
&+ \biggl(\frac{\Gamma_1 }{\Gamma_0 } - \frac{\beta_1}{\beta_0}\biggr) (1-r+\ln r)
   + \frac{\beta_1}{2\beta_0} \ln^2 r
\biggr\}
\,, \nn\\
\eta_\Gamma(\mu_0, \mu) &=
 - \frac{\Gamma_0}{2\beta_0}\, \biggl[ \ln r
 + \frac{\alpha_s(\mu_0)}{4\pi}\, \biggl(\frac{\Gamma_1 }{\Gamma_0 }
 - \frac{\beta_1}{\beta_0}\biggr)(r-1)
    \biggr]
\,, \nn\\
K_\gamma(\mu_0, \mu) &=
 - \frac{\gamma_0}{2\beta_0}\, \ln r
\, ,
\end{align}
with $r = \alpha_s(\mu)/\alpha_s(\mu_0)$. 

\subsection{Fixed-order hard matching functions}

The hard matching functions ${\cal H}_{c\to i}(z, p_T R, \mu)$ encode in them any radiation with virtuality of the order of ${\cal O}(p_T R)$, i.e., radiation that fails to pass the jet algorithm constraint. Up to NLO, it has been shown that they can be obtained by the out-of-jet radiation diagrams for the inclusive jet substructure observables~\cite{Kang:2016ehg}. 

From the knowledge of the one-loop structure of the hard matching functions, we can compute the RG equation which takes a DGLAP-like form. In order to solve it, we further decompose the complete one-loop matching function as
\begin{equation}
\mathcal{H}_{c\rightarrow i}(z,p_T R,\mu)=\mathcal{E}_c(p_T R,\mu)\,\mathcal{C}_{c\rightarrow i}(z,p_T R,\mu)\, .
\label{eq:matchingdecomp}
\end{equation}
Here, $\mathcal{E}_c$ is the non-DGLAP piece and satisfies an RG equation of the form
\begin{equation}
\mu \frac{d}{d\mu}\ln\mathcal{E}_c(\mu,p_TR)=\gamma_{c}(\mu,p_TR) \, ,
\label{eq:Ei}
\end{equation}
with the anomalous dimensions $\gamma_c$ given as 
\begin{equation}
\gamma_c =
\begin{cases}
\frac{\alpha_s(\mu) C_F}{\pi}\left(-L_h-\frac{3}{2}\right)&\quad : \quad \text{for quark jet}\, , \\
\frac{\alpha_s(\mu)}{\pi}\left(-C_A L_h-\frac{\beta_0}{2 }\right)& \quad : \quad \text{for gluon jet}\, .
\end{cases}
\end{equation}

The logarithm $L_h$ appearing in the anomalous dimensions is defined as
\begin{equation}
L_h = 2\ln\frac{\mu}{p_T R}\, .
\end{equation}
The function $\mathcal{C}_{c\rightarrow i}$ satisfies a DGLAP evolution equation which reads as

\begin{align}
&\frac{d}{d \ln \mu^2}\!
\begin{pmatrix}
	\mathcal{C}_{q\rightarrow g}(\mu,p_TR,z) \\ \mathcal{C}_{g\rightarrow g}(\mu,p_TR,z)
\end{pmatrix}
\!=\!
\frac{\alpha_s(\mu)}{4 \pi}\!\int_z^1\frac{dz'}{z'}\nonumber \\
&
\begin{pmatrix}
	\mathcal{P}_{gq}(z/z') & ~2 N_f \mathcal{P}_{q\bar{q}}(z/z') \\
	\mathcal{P}_{qg}(z/z') & ~\mathcal{P}_{gg}(z/z')
\end{pmatrix}
\otimes 
\begin{pmatrix}
	\mathcal{C}_{q\rightarrow g}(\mu,p_TR,z')\\ \mathcal{C}_{g\rightarrow g}(\mu,p_TR,z')
\end{pmatrix}\!.
\label{eq:dglap}
\end{align}

The solution of the hard functions from \eqs{eq:Ei}{eq:dglap} when combined with the jet and soft functions allows for the resummation of large logarithms of the measurement observable, leaving only logarithms of the jet radius $R$. These large logarithms can then resummed by evolving the result to the hard function scale. We provide the solution of the RG equations corresponding to the matching function in the section below.

\subsection{Renormalization Group evolution for matching functions}
\label{sec:matching}

The function ${\cal E}_c$ in \eq{eq:matchingdecomp} exhibits a simple multiplicative RG equation which has the solution
\begin{equation}
{\cal E}_c(\mu,p_TR)= e^{-K_{\Gamma_{\cal H}}^i-K_{\gamma_{\cal H}}^i} {\cal E}_c(\mu_{\cal H},p_TR)\, ,
\end{equation}
where $K_{\Gamma_{\cal H}}^i$, $K_{\gamma_{\cal H}}^i$ are evolution kernels defined as in \eqs{eq:Keta_def}{eq:Keta} and the index $i = \{ q, g\}$ is not summed over. For the hard function, the anomalous dimension coefficients are
\begin{align}
    &\Gamma_0^H = 4\, C_i\, , \,\, \Gamma_1^H = 4\, C_i \left[\left(\frac{67}{9}-\frac{\pi^2}{3}\right)C_A-\frac{20}{9}T_R N_f\right] \, ,\\
    &\gamma_0^{H, q} = 6\, C_F \, \, , \, \, \gamma_0^{H, g} = 2\,\beta_0\, .
\end{align}
The function $\cal C$ exhibits a standard DGLAP structure (see \eq{eq:dglap}). In order to solve \eq{eq:dglap}, we take the Mellin transform of which the $n^{th}$ moment is defined as
\begin{equation}
f(n)=\int_0^1 dx \, x^{n-1}f(x).
\end{equation}
This simplifies the convolution in \eq{eq:dglap} into a simple product, giving multiplicative coupled RG equations. Solving these coupled differential equations in Mellin's space using the eigenvalue method one obtains

\begin{align}
&\begin{pmatrix}
		\mathcal{C}_{q\rightarrow g}(\mu,p_TR,z) \\ \mathcal{C}_{g\rightarrow g}(\mu,p_TR,z)
	\end{pmatrix}
	=
\Big(e_{+}(n)\Big(\frac{\alpha_s(\mu)}{\alpha_s(\mu_{\cal H})}\Big)^{\!\!-r_{-}(n)}\!\!\!+e_{-}(n) \nonumber \\
&\qquad \times \Big(\frac{\alpha_s(\mu)}{\alpha_s(\mu_{\cal H})}\Big)^{\!\!-r_{+}(n)}\Big)	
	\begin{pmatrix}
		\mathcal{C}_{q\rightarrow g}(\mu_{\cal H},p_TR,z')\\ \mathcal{C}_{g\rightarrow g}(\mu_{\cal H},p_TR,z')
	\end{pmatrix}\!,
	\label{eq:match}
\end{align}

where the eigenvalues $r_{\pm}$ are given as

\begin{align}
r_{\pm}(n)&=\frac{1}{4\beta_0}\bigg(\mathcal{P}_{gq}(n)+\mathcal{P}_{gg}(n)\pm  \nonumber \\
& \sqrt{(\mathcal{P}_{gq}(n)-\mathcal{P}_{gg}(n))^2+8 N_f \mathcal{P}_{qg}(n)\mathcal{P}_{q\bar{q}}(n)}\bigg) \,  .
\end{align}

The eigenvectors $e_{\pm}$ in \eq{eq:match} read as

\begin{widetext}
\begin{equation}
e_{\pm}(n)=\frac{1}{r_{\pm}(n)-r_{\mp}(n)}\begin{pmatrix}
	\mathcal{P}_{gq}(n)/2-r_{\mp}(n) & ~ N_f \mathcal{P}_{q\bar{q}}(n) \\
	\mathcal{P}_{qg}(n)/2 & ~\mathcal{P}_{gg}(n)/2-r_{\mp}(n) 
\end{pmatrix} \, .
\end{equation}
\end{widetext}

Finally, the initial conditions for ${\cal C}$ required to obtain the complete solution of \eq{eq:match} are set by the explicit one-loop structure of the matching function.
This gives the complete solution of hard matching functions in the Mellin space. In order to obtain the solution in the physical space, we take the inverse Mellin transform of the solution obtained above numerically~\cite{Vogt:2004ns}.

\bibliographystyle{apsrev4-1}
\bibliography{main}
\end{document}